\documentclass[aps,prd,onecolumn,10pt,longbibliography,nofootinbib]{revtex4-2}

\usepackage{bold-extra}
\usepackage{anyfontsize}
\usepackage{comment}
\usepackage{calligra}
\usepackage{multirow}
\usepackage{physics}
\usepackage{amssymb}
\usepackage{graphicx}
\usepackage{booktabs}
\usepackage{nicefrac}
\usepackage{amsmath}
\usepackage{amsfonts}
\usepackage{array}
\usepackage{mathrsfs}
\usepackage[mathscr]{euscript}
\usepackage{stfloats}
\usepackage{bm}
\usepackage{float}

\usepackage{hyperref}
\hypersetup{
    colorlinks=true,
    linkcolor=blue,
    filecolor=magenta,      
    urlcolor=cyan,
    citecolor=red,
}
\usepackage[dvipsnames]{xcolor}
\usepackage[normalem]{ulem}
\newcommand{\modecasecustom}[2]{%
    \medskip
    \noindent\textbf{\textit{(#1) #2}}\par
    \smallskip
}
\begin{document}
\setlength{\parindent}{1.5em}
\setlength{\parskip}{0.6em}

\title{Another Look at the Weak-Field Limit of Generalized Hybrid Metric-Palatini Gravity}
\author{Gustavo Melgarejo}
\email{ga.melgarejoc@gmail.com}
\affiliation{Universidade do Estado do Rio de Janeiro (UERJ), CEP 20550-013, Rio de Janeiro, RJ, Brazil}

\author{Santiago Esteban Perez Bergliaffa}
\email{sepbergliaffa@gmail.com}
\affiliation{Universidade do Estado do Rio de Janeiro (UERJ), CEP 20550-013, Rio de Janeiro, RJ, Brazil}
\newcommand{\santiago}[1]{\textcolor{orange}{#1}}
\newcommand{\gustavo}[1]{\textcolor{cyan}{#1}}


\begin{abstract}

We investigate the weak-field regime of generalized hybrid metric-Palatini theories, described by a generic function \(f(R,\mathcal{R})\), where \(R\) is the metric Ricci scalar and \(\mathcal{R}\) is constructed from an independent torsionless connection. Linearizing the field equations about Minkowski spacetime, we show, without using the scalar-tensor representation, that the theory propagates the usual massless spin-2 mode and two massive scalar modes, with an effective gravitational coupling. The absence of tachyonic and ghostlike instabilities at the linearized level, together with the nondegeneracy of the scalar sector, is shown to impose algebraic restrictions on the derivatives of \(f(R,\mathcal R)\) evaluated on the Minkowski background, which generalize previously obtained conditions. The Newtonian limit for an extended static source is derived, yielding a gravitational potential with two Yukawa corrections whose amplitudes are fixed by the scalar residues, while finite-size effects are encoded in source-dependent form factors. We determine the conditions under which the usual Newtonian limit is recovered and derive the effective post-Newtonian parameter \(\gamma_\Sigma\) governing light propagation. Finally, we compute the radial epicyclic frequency and the corresponding anomalous periapsis advance, and compare it with planetary precession data to constrain the parameters of a viable hierarchical scalar-mass regime.
\end{abstract}

\maketitle


\section{ Introduction}

The accelerated expansion of the universe \cite{Ries1998,Perlmutter1999} calls for a critical examination of the pillars of the standard cosmological model. In particular, in view of the possible time variation of such a phenomenon \cite{DESI2025}, it is natural to consider 
theories of gravity that differ from General Relativity (GR) \cite{Clifton2011,Nojiri2017}. The possible ways in which GR can be modified are classified following Lovelock’s theorem in
\cite{Berti2015}. Among
the many options, 
geometrical modifications of the Einstein-Hilbert Lagrangian, such as
$f(R)$ theories,
have been studied both in the metric and in the Palatini version
\cite{Sotiriou2008,DeFelice2010}. Since these proposals are not free of problems (see for instance \cite{Ishak2018, Ventagli2023}), they were further generalized to the so-called hybrid theories, the linear case of which, first introduced in 
\cite{Harko2011}, is defined by $\mathcal{L}=R+f(\mathcal{R})$, 
where 
$R$ is the Ricci scalar built using the Christoffel symbols, and 
$\mathcal{R}$ is the Ricci scalar built using a connection which is \textit{a priori} independent of the metric.
Such a theory, which includes a scalar degree of freedom as well as the metric,
has been examined in several scenarios 
(see \cite{Capo2015} for a review). The so-called generalized  
hybrid theories 
(GHT)
are those in which the Lagrangian is an arbitrary function of both $R$ and $\mathcal{R}$
\cite{Flanagan2003,PhysRevD.87.084031}. Besides the metric, these theories, which admit a scalar-tensor representation \cite{Flanagan2003}, display two scalar degrees of freedom \cite{Flanagan2003}, 
which could be used to source the accelerated expansion and describe dark matter (see 
\cite{Bombacigno,Dyadina2025} for recent developments, and
\cite{vanderbruck2023} and references therein).

Although several aspects of the GHT have been explored in the literature 
(see a list of references in \cite{Gomes2025}), 
only a few papers deal with 
constraining the huge freedom inherent to the
functional form of $f(R,\mathcal{R})$
\footnote{See for instance \cite{Rosa2020,Rosa2024}.}
. Such a goal can be achieved in particular in the weak-field limit, a method that has proven to be useful in $f(R)$ theories \cite{Zakharov2006,Olmo2005}, and in the 
linear case of hybrid theories
\cite{Capo2015}.
The weak-field limit of  GHT has been studied 
in \cite{Bombacigno2019}, where the scalar-tensor representation in the Jordan frame was used, and in 
\cite{Rosa2021}, in the Einstein frame. In both cases, the modifications to the Newtonian potential and the parameter $\gamma$ for point-like masses were obtained.
Here we shall reexamine the weak-field limit by carrying out the calculation 
without using the scalar-tensor representation, in what could be called the original representation of GHT, in terms of the variables $R$ and $\mathcal{R}$, and  considering a general source. By perturbing the metric tensor with respect to the Minkowskian background, we shall show how the perturbations can be decomposed into a spin 2 mode and two scalar degrees of freedom, each obeying its own equation of motion. 
Our calculations generalize those presented in 
\cite{Koivisto2013}, since we do not assume that
$\frac{d^2f}{d\mathcal{R}^2}=0$ in the background.
We shall also obtain the constraints that follow from imposing that the scalar degrees of freedom do not behave as ghosts or tachyons, determine the conditions under which the Newtonian limit is recovered, and apply 
the results of the weak-field limit to constrain the relevant parameters using data from epicyclic radial frequencies.

Let us point out that the weak field limit, including Yukawa corrections, an effective Newtonian coupling, a post-Newtonian parameter, and Solar-System phenomenology, has been recently studied in \cite{Pereira2025} in the case of a sub-class of an extended $f(R,\mathcal{R})$ theory which incorporates an extra scalar. The calculations were carried out in the Jordan frame, for a specific form of the Lagrangian.

The structure of the paper is the following. In Sect. \ref{actionfrr}, the general features of GHT will be presented. Section \ref{wfl} shows how the perturbation of the equations of motion of GHT with respect to the Minkowskian background can be decomposed into a tensor mode and two scalar modes.
In Section \ref{constraints}, the propagator of the theory is calculated, and the constraints that follow from imposing the absence of ghosts and tachyons are deduced. 
The weak-field limit leading to the modified Newtonian potential and the $\gamma$ parameter
are presented in Sect.\ref{section 5}. The resulting expressions are then analyzed for different types of sources, and different relations between the masses of the scalar degrees of freedom. We present a discussion of the appropriate form of the parameter $\gamma$
in the Appendix \ref{appendix}.
Section \ref{section 5} also includes the limits on the parameters of GHT
appearing in the weak-field limit when applied to epicyclic radial frequencies. We close in Sect.\ref{concl} with some comments on the results and future perspectives.

Throughout this work we adopt the conventions $\kappa=8\pi G$, $c=\hbar=1$, and the metric signature $(-1,+1,+1,+1)$. Here $G$ denotes the bare gravitational constant appearing in the action and is not identified a priori with the Newtonian constant inferred from observations.

\section{Action for \texorpdfstring
{$f(R,\mathcal{R})$}~ gravity  and field equations}
\label{actionfrr}

The action for generalized hybrid metric-Palatini gravity is given by
\begin{equation}
    S=\frac{1}{2\kappa}\int d^4x\sqrt{-g}~f\left(R,\mathcal{R}\right)+S_m,
    \label{actionf}
\end{equation}
where $S_m$ is the matter action, minimally coupled to $g_{\mu\nu}$, and indices are raised and lowered with $g_{\mu\nu}$. The metric Ricci scalar is defined as $R=g^{\mu\nu}R_{\mu\nu}$, where
\begin{equation}
    R_{\mu\nu}
    =\partial_{\lambda}\Gamma^{\lambda}{}_{\mu\nu}
     -\partial_{\nu}\Gamma^{\lambda}{}_{\mu\lambda}
     +\Gamma^{\sigma}{}_{\mu\nu}\Gamma^{\lambda}{}_{\sigma\lambda}
     -\Gamma^{\sigma}{}_{\mu\lambda}\Gamma^{\lambda}{}_{\sigma\nu},
\end{equation}
is the Ricci tensor constructed from the Levi--Civita connection $\Gamma^{\rho}{}_{\mu\nu}$ of the metric. The Palatini Ricci scalar is defined by $\mathcal{R}=g^{\mu\nu}\mathcal{R}_{\mu\nu}=g^{\mu\nu}\mathcal{R}_{(\mu\nu)}$, where $\mathcal{R}_{\mu\nu}$ is the Ricci tensor associated with an independent connection $\widetilde{\Gamma}^{\rho}{}_{\mu\nu}$
\begin{equation}
    \mathcal{R}_{\mu\nu}
    =\partial_{\lambda}\widetilde{\Gamma}^{\lambda}{}_{\mu\nu}
     -\partial_{\nu}\widetilde{\Gamma}^{\lambda}{}_{\mu\lambda}
     +\widetilde{\Gamma}^{\sigma}{}_{\mu\nu}\widetilde{\Gamma}^{\lambda}{}_{\sigma\lambda}
     -\widetilde{\Gamma}^{\sigma}{}_{\mu\lambda}\widetilde{\Gamma}^{\lambda}{}_{\sigma\nu}.
\end{equation}

Variation of the action \eqref{actionf} with respect to the metric yields the field equations
\begin{equation}
    f_{,R}R_{\mu\nu}-\frac{1}{2}g_{\mu\nu}f-\left(\nabla_{\mu}\nabla_{\nu}-g_{\mu\nu}\Box_g\right)f_{,R}+f_{,\mathcal{R}}\mathcal{R}_{(\mu\nu)}=\kappa\,T_{\mu\nu},
    \label{field-equation}
\end{equation}
where $f_{,R}$ and $f_{,\mathcal{R}}$ denote the partial derivatives of $f$ with respect to $R$ and $\mathcal{R}$, respectively, and $\Box_g\equiv g^{\mu\nu}\nabla_{\mu}\nabla_{\nu}$. The matter stress-energy tensor is defined as
\begin{equation}
    T_{\mu\nu} \equiv -\frac{2}{\sqrt{-g}} 
    \frac{\delta S_m}{\delta g^{\mu\nu}} .
\end{equation}

Variation of \eqref{actionf} with respect to the independent connection $\widetilde{\Gamma}^\rho_{\mu\nu}$ gives
\begin{equation}
     \tfrac{1}{2\sqrt{-g}}\widetilde{\nabla}_{\alpha}(\sqrt{-g}q^{\mu\nu})=q^{\mu\sigma}S^{\nu}{}_{\alpha\sigma}+q^{\mu\nu}S^{\sigma}{}_{\sigma\alpha}-\frac{1}{3}\delta^{\nu}_{\alpha}q^{\mu\rho}S^{\sigma}{}_{\sigma\rho},    \label{eq:83}
\end{equation}
where
\begin{equation}
    q^{\mu\nu}\equiv
    f_{,\mathcal{R}}g^{\mu\nu},
    \label{eq:q}
\end{equation}
$\widetilde{\nabla}_{\alpha}$ denotes the covariant derivative associated with $\widetilde{\Gamma}^\rho{}_{\mu\nu}$, and $S^\lambda{}_{\mu\nu}\equiv\widetilde{\Gamma}^\lambda{}_{[\mu\nu]}$ is the torsion tensor. 
A common simplification, which we shall adopt here, is to assume a torsionless connection. 
Hence, Eq.~\eqref{eq:83} reduces to
\begin{equation}
     \widetilde{\nabla}_{\alpha}(\sqrt{-g}q^{\mu\nu})=0.   \label{eq:84}
\end{equation}
To solve Eq.~\eqref{eq:84}, it is convenient to introduce an auxiliary metric 
$\tilde{g}_{\mu\nu}$ defined by
\begin{equation}
\label{Gamma-ind}
\sqrt{-\tilde{g}}\,\tilde{g}^{\mu\nu}=\sqrt{-g}\, q^{\mu\nu}.
\end{equation}
For any GHT, $f_{,\mathcal{R}}$ depends implicitly on $\widetilde{\Gamma}^\mu{}_{\nu\lambda}$ through \(\mathcal{R}\), so a closed algebraic expression for 
{$\widetilde{\Gamma}^\mu{}_{\nu\lambda}$}
cannot be obtained. However, the following expression follows from 
Eq.\eqref{Gamma-ind}:
\begin{equation}
    \widetilde{\Gamma}_{\nu \lambda}^\mu=\frac{1}{2}\tilde{g}^{\mu\rho} (\partial_\nu \tilde{g}_{\rho \lambda}+\partial_\lambda \tilde{g}_{\rho \nu}-\partial_\rho \tilde{g}_{\nu \lambda} ) \label{gamma-palatini}.
\end{equation}
 As we will show later, in the weak-field limit the right-hand side of Eq.~\eqref{gamma-palatini} becomes independent of $\widetilde{\Gamma}^\mu{}_{\nu\lambda}$ at first perturbative order.
Using Eq.~\eqref{gamma-palatini}, the Palatini Ricci tensor can be written as
\begin{equation}
\mathcal{R}_{(\mu\nu)}=R_{\mu\nu}+\frac{3}{2(f_{,\mathcal{R}})^2}\partial_\mu f_{,\mathcal{R}} \partial_\nu f_{,\mathcal{R}}-\frac{1}{ f_{,\mathcal{R}}}\left(\nabla_{\mu}\nabla_{\nu}+\frac{1}{2}g_{\mu\nu}\Box_g\right)f_{,\mathcal{R}}.
\label{ricci-calig}
\end{equation}
Taking the trace of Eqs.~\eqref{field-equation} and \eqref{ricci-calig}, one obtains the following coupled differential equations for the curvature scalars:
\begin{equation}
\begin{aligned}
  &3\Box_g f_{,R}+ f_{,R} R+f_{,\mathcal{R}}\mathcal{R}-2f=\kappa\,T, \\&3\Box_g f_{,\mathcal{R}}- f_{,\mathcal{R}} R+ f_{,\mathcal{R}}\mathcal{R}-\frac{3}{2f_{,\mathcal{R}}}\partial_\mu f_{,\mathcal{R}} \partial^\mu f_{,\mathcal{R}}=0. 
\end{aligned}
\label{dof-equations}
\end{equation}
Equations \eqref{field-equation} and \eqref{dof-equations} therefore form a highly coupled nonlinear system for the metric and the curvature scalars. It also follows from these equations that, in general, GHT propagates two scalar degrees of freedom in addition to the metric tensor.

\section{Weak-field limit and physical modes}
\label{wfl}

In this section, we study the weak-field limit of GHT by linearizing the field equations around Minkowski spacetime\footnote{We follow the method of \cite{Koivisto2013} for the perturbative equations and the propagator, while allowing $f^{(0)}_{,\mathcal{R}\mathcal{R}}\neq 0$.}. This analysis identifies the masses of the propagating degrees of freedom and prepares the derivation of the modified Newtonian potential and the so-called $\gamma$ parameter. 

Let us start by considering  small perturbations around Minkowski spacetime, so that the metric can be written as
\begin{equation}
    g_{\mu\nu}= \eta_{\mu\nu}+h_{\mu\nu},
    \label{lin-metric}
\end{equation}
where $\eta_{\mu\nu}$ is the Minkowski metric and $|h_{\mu\nu}|\ll1$. 

Up to first order in $h_{\mu\nu}$, one has
\begin{equation}
    g^{\mu\nu}\approx \eta^{\mu\nu}-h^{\mu\nu},
    \label{lin-invermet}
\end{equation}
where indices are raised and lowered with $\eta_{\mu\nu}$, e.g., $h^{\mu\nu}=\eta^{\mu\sigma}\eta^{\nu\rho}h_{\sigma\rho}$. The Levi--Civita connection of the perturbed metric can then be expanded as
\begin{equation}
    \Gamma_{~\nu \lambda}^{\mu}=\frac{1}{2} \eta^{\mu\rho} (\partial_\nu h_{\rho\lambda}+\partial_\lambda h_{\rho \nu}-\partial_\rho h_{\nu \lambda} )+\mathcal{O}(h^2),
\end{equation}
so that, at this order, covariant derivatives reduce to partial derivatives. The linearized Ricci tensor is therefore
\begin{equation}
R_{\mu\nu}=\partial_\sigma\partial_{(\mu} h_{~ \nu)}^\sigma-\frac{1}{2}\Box h_{\mu\nu}-\frac{1}{2}\partial_\mu\partial_\nu h+\mathcal{O}(h^2),
\end{equation}
where $\Box\equiv\eta^{\mu\nu}\partial_\mu\partial_\nu$ and $h\equiv\eta^{\mu\nu}h_{\mu\nu}$.
\\
The expansion of $f(R,\mathcal{R})$ around the Minkowski background takes the form
\begin{equation}
    f (R,\mathcal{R})=f^{(0)}+f^{(0)}_{,R} R^{(1)}+f^{(0)}_{,\mathcal{R}} \mathcal{R}^{(1)}+\mathcal{O}(h^2),
\end{equation}
with similar expansions for its derivatives. Here $f^{(0)}$, $f^{(0)}_{,R}$, and $f^{(0)}_{,\mathcal{R}}$ denote constant values evaluated at $R=\mathcal{R}=0$.
The quantities $R^{(1)}$ and $\mathcal{R}^{(1)}$ represent the first-order contributions to $R$ and $\mathcal{R}$, respectively.

From Eq.~\eqref{Gamma-ind} it follows that $\sqrt{-\tilde{g}}=(\sqrt{-g})^2/\sqrt{-q}$. Consequently,
\begin{equation}
    \tilde{g}^{\mu\nu}
    =\frac{\sqrt{-q}}{\sqrt{-g}}\,q^{\mu\nu}=\frac{1}{f_{,\mathcal{R}}}g^{\mu\nu},
    \label{eq:conf-metric}
\end{equation}
with $g=\det(g_{\mu\nu})$ and $q\equiv \det(q_{\mu\nu})$, where $q_{\mu\nu}$ is the inverse matrix of $q^{\mu\nu}$.
\\
Since \(\tilde{g}_{\mu\nu}=f_{,\mathcal{R}}\,g_{\mu\nu}\), the perturbed conformal metric can be written as
\begin{equation}
    \tilde{g}_{\mu\nu}
    = f^{(0)}_{,\mathcal{R}}\left(\eta_{\mu\nu}+\tilde{h}_{\mu\nu}\right)
    + \mathcal{O}(h^2),
    \label{g-down}
\end{equation}
where
\begin{equation}
    \tilde{h}_{\mu\nu}
    = h_{\mu\nu}
    + \eta_{\mu\nu}\left(
    \mathcal{A}\,R^{(1)}
    + \mathcal{B}\,\mathcal{R}^{(1)}
    \right),
\end{equation}
with
 $$   \mathcal{A}\equiv\frac{f^{(0)}_{,R\mathcal{R}}}{f^{(0)}_{,\mathcal{R}}},
    \qquad
    \mathcal{B}\equiv\frac{f^{(0)}_{,\mathcal{R}\mathcal{R}}}{f^{(0)}_{,\mathcal{R}}}.
$$

Using Eqs.~\eqref{eq:conf-metric} and \eqref{g-down}, the independent connection \(\widetilde{\Gamma}^{\rho}{}_{\mu\nu}\) can be expanded to first order in $h_{\mu\nu}$ as
\begin{equation}
    \widetilde{\Gamma}_{\nu \lambda}^\mu=\frac{1}{2}\eta^{\mu\rho} (\partial_\nu \tilde{h}_{\rho \lambda}+\partial_\lambda \tilde{h}_{\rho \nu}-\partial_\rho \tilde{h}_{\nu \lambda} )+\mathcal{O}(h^2).
\end{equation}
It then follows that the Palatini Ricci tensor can be written as
\begin{equation}
    \mathcal{R}_{(\mu\nu)}=R^{(1)}_{\mu\nu}-\mathcal{A}\,\partial_{\mu}\partial_{\nu}R^{(1)}-\mathcal{B}\,\partial_{\mu}\partial_{\nu}\mathcal{R}^{(1)}-\frac{1}{2}\mathcal{A}\,\eta_{\mu\nu}\,\Box R^{(1)}-\frac{1}{2}\mathcal{B}\,\eta_{\mu\nu}\,\Box\mathcal{R}^{(1)}+\mathcal{O}(h^2).
    \label{eq:Ricci-cal}
\end{equation}

Substituting these expressions into the field equations \eqref{field-equation}, we obtain the linearized equations
\begin{eqnarray}
\nonumber&&f_{,R}^{(0)} R^{(1)}_{\mu\nu}-\frac{1}{2}\eta_{\mu\nu}\left(f_{,R}^{(0)} R^{(1)}+ f_{,\mathcal{R}}^{(0)}\mathcal{R}^{(1)}\right)+f^{(0)}_{,RR}\,\eta_{\mu\nu}\,\Box R^{(1)} -\mathcal{A}\,f_{,\mathcal{R}}^{(0)}\,\partial_\mu\partial_\nu \mathcal{R}^{(1)}+\\ && -f^{(0)}_{,RR}\,\partial_\mu\partial_\nu R^{(1)}
 +\mathcal{A}\,f_{,\mathcal{R}}^{(0)}\,\eta_{\mu\nu}\,\Box \mathcal{R}^{(1)}+f_{,\mathcal{R}}^{(0)}\mathcal{R}^{(1)}_{(\mu\nu)}
-\frac{1}{2} f^{(0)}h_{\mu\nu}= \kappa\,T^{(1)}_{\mu\nu},
\label{mov-eq-firstorder}
\end{eqnarray}
where $T^{(1)}_{\mu\nu}$ denotes the linear perturbation of the stress-energy tensor. 

The last term in Eq.~\eqref{mov-eq-firstorder} is proportional to the background value $f^{(0)}$, which plays the role of an effective cosmological constant. Since the perturbative expansion is performed around Minkowski spacetime, {the condition
$f^{(0)}=0$ must be imposed}.

Taking the trace of Eqs.~\eqref{eq:Ricci-cal} and \eqref{mov-eq-firstorder}, and assuming $\mathcal{A}+\mathcal{B}\neq 0$, yields
\begin{equation}
\mathcal{R}^{(1)}
=\frac{1} {f_{,\mathcal{R}}^{(0)}(\mathcal{A}+\mathcal{B})}\Bigg[\Big(\mathcal{A}\,f_{,\mathcal{R}}^{(0)}-\mathcal{B}\,f_{,R}^{(0)}\Big)\,R^{(1)}+3\Big(\mathcal{B}\,f_{,RR}^{(0)}-\mathcal{A}^2\,f_{,\mathcal{R}}^{(0)}\Big)\,\Box R^{(1)}
-\mathcal{B}\,\kappa\, T^{(1)}\Bigg].
\label{R-caligrafico}
\end{equation}
Equivalently, Eq.~\eqref{R-caligrafico} also follows from the first-order
expansion of Eqs.~\eqref{dof-equations}.

Using Eq.\eqref{eq:Ricci-cal}, 
we can write Eq.~\eqref{mov-eq-firstorder} as follows:
\begin{eqnarray}\label{29}
\nonumber&&(f_{,R}^{(0)}+f_{,\mathcal{R}}^{(0)})\,R^{(1)}_{\mu\nu}-\Big(\mathcal{A}\,f_{,\mathcal{R}}^{(0)}+f_{,RR}^{(0)}\Big)\,\partial_\mu\partial_\nu R^{(1)}
-f_{,\mathcal{R}}^{(0)}\,\big(\mathcal{A}+\mathcal{B}\big)\,\partial_\mu\partial_\nu \mathcal{R}^{(1)}
-\frac{1}{2}\,f_{,R}^{(0)}\,\eta_{\mu\nu} R^{(1)}
-\frac{1}{2}\,f_{,\mathcal{R}}^{(0)}\,\eta_{\mu\nu} \mathcal{R}^{(1)}+
\\[4pt]&&+\frac{1}{2}\Big(2f_{,RR}^{(0)}-\mathcal{A}\,f_{,\mathcal{R}}^{(0)}\Big)\,\eta_{\mu\nu}\Box R^{(1)}+\frac{1}{2}f_{,\mathcal{R}}^{(0)}\big(2\mathcal{A}-\mathcal{B}\big)\,\eta_{\mu\nu}\,\Box\mathcal{R}^{(1)}
=\kappa\,T^{(1)}_{\mu\nu}.
\end{eqnarray}
$\mathcal{R}^{(1)}$ 
can be replaced in the latter equation 
by Eq.~\eqref{R-caligrafico}. Using the trace of the resulting equation to simplify the remaining terms, Eq.~\eqref{29} can be recast as
\begin{equation}
2G^{(1)}_{\mu\nu}
+\left[1-\frac{c(\Box)}{a}\right]\eta_{\mu\nu}R^{(1)}-\frac{e(\Box)}{a}\partial_\mu\partial_\nu R^{(1)}=2\kappa_{\mathrm{eff}} T^{\mathrm{eff}}_{\mu\nu},
\label{eq:O-in-GR}
\end{equation}  
where
\begin{equation*}
   T^{\mathrm{eff}}_{\mu\nu}= T_{\mu\nu}^{(1)}- \mathcal{B}\partial_{\mu}\partial_{\nu}T^{(1)} + \mathcal{B}\eta_{\mu\nu}\Box  T^{(1)},
\end{equation*}
represents an effective linearized source, and
 $$   G^{(1)}_{\mu\nu}\equiv R^{(1)}_{\mu\nu}-\frac{1}{2}\eta_{\mu\nu}R^{(1)},
$$
is the linearized Einstein tensor, with
 $   \kappa_{\mathrm{eff}}\equiv\frac{\kappa}{a}
 $.
The quantities $a$, $c(\Box)$, and $e(\Box)$ are given by
\begin{align}
    a &= \left(f_{,R}^{(0)}+f_{,\mathcal{R}}^{(0)}\right),
    \label{a-cuadrito}
\\
    c(\Box) &= \left(f_{,R}^{(0)}+f_{,\mathcal{R}}^{(0)}\right) 
- 2\left(2\mathcal{A}\,f_{,\mathcal{R}}^{(0)}+f_{,RR}^{(0)}-\mathcal{B}\,f_{,R}^{(0)}\right)\Box
+ 6\left(\mathcal{A}^2\,f_{,\mathcal{R}}^{(0)}
-\mathcal{B}f_{,RR}^{(0)}\right)\Box^2,
\label{c-cuadrito}
\\
    e(\Box) &= 2\left(2\mathcal{A}\,f_{,\mathcal{R}}^{(0)}+f_{,RR}^{(0)}-\mathcal{B}\,f_{,R}^{(0)}\right)
- 6\left(\mathcal{A}^2\,f_{,\mathcal{R}}^{(0)}
- \mathcal{B}\,f_{,RR}^{(0)}\right)\Box.
\label{e-cuadrito}
\end{align}
Since the linearized Einstein tensor satisfies the contracted Bianchi identity, \(\partial^{\mu}G^{(1)}_{\mu\nu}=0\), and since \(c(\Box)+\Box\,e(\Box)-a=0\), one verifies by direct differentiation that \(\partial^{\mu}T^{(1)}_{\mu\nu}=0\) identically. The linearized field equations are therefore compatible with the Bianchi identity.

In the GR limit, one has
\[
f^{(0)}_{,\mathcal{R}\mathcal{R}} = f^{(0)}_{,R\mathcal{R}} = f^{(0)}_{,RR} = 0,
\qquad
f^{(0)}_{,R} + f^{(0)}_{,\mathcal{R}} = 1,
\]
so that the expected results
\[
a=c(\Box)=1,
\qquad
e(\Box)=0,
\]
are recovered.

To identify the scalar modes explicitly, we take the trace of Eq.~\eqref{eq:O-in-GR} and factorize the resulting operator. This yields
\begin{equation}
    \frac{2}{m_+^2 m_-^2}
    \big(\Box-m_+^2\big)\big(\Box-m_-^2\big)R^{(1)}
    = -2\kappa_{\mathrm{eff}} \left(1+3\mathcal{B}\,\Box\right)T^{(1)},
\end{equation}
where
\begin{equation}
    m_{\pm}^2 =
    \frac{
    2\mathcal{A}\,f_{,\mathcal{R}}^{(0)}
    - \mathcal{B}\,f_{,R}^{(0)}
    + f_{,RR}^{(0)}
    \pm \mathcal{S}
    }{
    6\left(\mathcal{A}^{2}f_{,\mathcal{R}}^{(0)}
    - \mathcal{B}\,f_{,RR}^{(0)}\right)
    },
    \label{masas-campos}
\end{equation}
with
\begin{equation}
    \mathcal{S}^2 = \left(\mathcal{B}\,f_{,R}^{(0)} - 2\mathcal{A}\,f_{,\mathcal{R}}^{(0)} - f_{,RR}^{(0)}\right)^2-4\left(f_{,\mathcal{R}}^{(0)} + f_{,R}^{(0)}\right)\left(\mathcal{A}^2\,f_{,\mathcal{R}}^{(0)}
- \mathcal{B}\,f_{,RR}^{(0)}\right).
\end{equation}
In the generic case \(\mathcal{S}\neq 0\),  the fourth-order equation for \(R^{(1)}\) can be decomposed into two second-order equations. To make this structure explicit, we introduce the field
\begin{equation}
   \phi \equiv \frac{2}{m_+^2 m_-^2}\left(\Box - m_-^2\right)R^{(1)},
   \label{phi}
\end{equation}
which satisfies
\begin{equation}
    \left(\Box - m_+^2\right)\phi = -2\kappa_{\mathrm{eff}} \left(1+3\mathcal{B}\,\Box\right)T^{(1)}.
    \label{eq:mov-phi}
\end{equation}
Similarly, defining
\begin{equation}
   \psi \equiv \frac{2(m_-^2 - m_+^2)}{m_+^2 m_-^2}R^{(1)} + \phi
   = \frac{2}{m_+^2 m_-^2}\left(\Box - m_+^2\right)R^{(1)},
   \label{psi}
\end{equation}
it follows that
\begin{equation}
    \left(\Box - m_-^2\right)\psi = -2\kappa_{\mathrm{eff}} \left(1+3\mathcal{B}\,\Box\right)T^{(1)}.
    \label{eq:mov-psi}
\end{equation}

Having established the factorized scalar structure, we now return to the original linearized field equations and recast them in a form analogous to the gravitational-wave equation. This makes the separation between the scalar and tensor sectors more transparent.

Equation \eqref{eq:O-in-GR} can be written in terms of $h_{\mu\nu}$ and its trace as
\begin{equation}
2\left(
\partial_\sigma \partial_{(\mu} h^\sigma{}_{\nu)}
-\frac{1}{2}\Box h_{\mu\nu}
-\frac{1}{2}\partial_\mu \partial_\nu h
\right)
-\frac{1}{a}\left[\eta_{\mu\nu}\,c(\Box)
+e(\Box)\,\partial_\mu \partial_\nu\right]
\left(
\partial_\alpha \partial_\beta h^{\alpha\beta}
-\Box h
\right)
=
2\kappa_{\mathrm{eff}} T^{\mathrm{eff}}_{\mu\nu}.
\label{expanded-field-eq-FO}
\end{equation}
Imposing the de Donder gauge,
\begin{equation}
\partial_\mu h^{\mu\nu}=\frac{1}{2}\partial^{\nu}h,
\end{equation}
equation~\eqref{expanded-field-eq-FO} therefore reduces to
\begin{equation}
\Box h_{\mu\nu}
-\frac{1}{2}\frac{c(\Box)}{a}\,\eta_{\mu\nu}\,\Box h
-\frac{1}{2}\frac{e(\Box)}{a}\,\partial_\mu\partial_\nu \Box h
=
-2\kappa_{\mathrm{eff}} T^{\mathrm{eff}}_{\mu\nu}.
\label{expanded-field-eq-FO-deDonder}
\end{equation}
This motivates the field redefinition
\begin{equation}
    \hat{h}_{\mu\nu}\equiv h_{\mu\nu}
-\frac{1}{2}\frac{c(\Box)}{a}\,\eta_{\mu\nu}\, h
-\frac{1}{2}\frac{e(\Box)}{a}\,\partial_\mu\partial_\nu h.
\label{hat-h}
\end{equation}
Using the relation \(c(\Box)+\Box\,e(\Box)-a=0\), it follows that
\begin{equation}
    \partial^{\mu}\hat{h}_{\mu\nu}=\partial^{\mu}h_{\mu\nu}-\frac{1}{2}\partial_{\nu}h,
\end{equation}
Thus, in the de Donder gauge, \(\hat{h}_{\mu\nu}\) is transverse
\begin{equation}
\partial_{\mu}\hat{h}^{\mu\nu}=0.
\end{equation}
Equation \eqref{expanded-field-eq-FO-deDonder} then becomes
\begin{equation}
      \Box \hat{h}_{\mu\nu} =-2\kappa_{\mathrm{eff}} T^{\mathrm{eff}}_{\mu\nu},
    \label{eq:mov-hat-h}
\end{equation}
showing that \(\hat{h}_{\mu\nu}\) satisfies the standard massless wave equation, sourced by \(T^{\mathrm{eff}}_{\mu\nu}\) with effective gravitational coupling \(\kappa_{\mathrm{eff}}\). It can therefore be identified with the massless spin-2 sector of the theory.

Using
\begin{equation}
    \left(2\mathcal{A}\,f_{,\mathcal{R}}^{(0)}+f_{,RR}^{(0)}-\mathcal{B}\,f_{,R}^{(0)}\right)=\frac{a}{3}\frac{m_+^2 +m_-^2}{m_+^2 m_-^2},\qquad\text{and}\qquad \left(\mathcal{A}^2\,f_{,\mathcal{R}}^{(0)}
- \mathcal{B}\,f_{,RR}^{(0)}\right)=\frac{a}{9 m_+^2 m_-^2},
\end{equation}
together with the definitions~\eqref{phi}, \eqref{psi}, and \eqref{hat-h}, one can express \(h\) and \(\Box h\) in terms of \(\phi\), \(\psi\), and \(\hat h\). The metric perturbation then decomposes as
\begin{eqnarray}
    \nonumber
    h_{\mu\nu}
    &=&\hat{h}_{\mu\nu}- \frac{1}{2}\eta_{\mu\nu}\hat{h}-\frac{m_+^2+m_-^2}{3m^2_+m^2_-}\partial_{\mu}\partial_{\nu}\hat{h}-\frac{1}{6}\frac{m^2_++m^2_-}{m^2_+ -m^2_-}\eta_{\mu\nu}(\phi-\psi)+\frac{1}{6}\frac{1}{m_+^2-m_-^2}\eta_{\mu\nu}\,\Box(\phi-\psi)+\\\nonumber&&-\frac{m_-^4+m_-^2 m_+^2+m_+^4}{3 m_+^2 m_-^2 \left(m_+^2-m_-^2\right)}\partial_{\mu}\partial_{\nu}(\phi-\psi)+\frac{m^2_+ + m^2_-}{3m^2_+ m^2_-(m^2_+ -m^2_-)}\,\partial_{\mu}\partial_{\nu}\Box(\phi-\psi),
    \label{h-grados-de-libertad}
\end{eqnarray}
where \(\phi\) and \(\psi\) denote scalar modes with masses \(m_+\) and \(m_-\), respectively.
This expression, 
{which satisfies Eq.\eqref{expanded-field-eq-FO-deDonder}},
shows
that the perturbation $h_{\mu\nu}$ can be obtained by solving the wave-like equations for $\hat h_{\mu\nu}$, $\phi$, and $\psi$, which have the first-order perturbation of the stress-energy tensor, and its trace as a source, respectively.

\subsection{Degenerate Cases}
\label{sec:degenerate}
The physical interpretation of the spectrum derived above relies on two generic assumptions: 
\begin{enumerate} 
    \item $\mathcal{S} \neq 0$, and
    \item $\mathcal{A}^2 f_{,\mathcal{R}}^{(0)} - \mathcal{B} f_{,RR}^{(0)} \neq 0$.
\end{enumerate}
When either of these conditions is violated, the structure of the spectrum may change, potentially altering the number and nature of the propagating degrees of freedom. The corresponding degenerate regimes are discussed next.

\paragraph{Coincident Masses ($\mathcal{S}^2 = 0$).}
When the discriminant vanishes, the two scalar masses coincide ($m_+^2 = m_-^2 \equiv m^2$). In this limit, the independent field definitions for $\phi$ and $\psi$ in Eqs.~\eqref{phi} and \eqref{psi} 
yield the same expression, while the metric reconstruction in Eq.~\eqref{h-grados-de-libertad} diverges due to the vanishing denominator $m_+^2 - m_-^2$. Furthermore, the fourth-order operator reduces to a repeated root $(\Box - m^2)^2$. Consequently, the standard two-field decoupling procedure is no longer applicable, and these degenerate cases must be analyzed via a separate, dedicated treatment of the field equations.

\paragraph{Loss of Higher-Derivative Dynamics ($\mathcal{A}^2 f_{,\mathcal{R}}^{(0)} - \mathcal{B} f_{,RR}^{(0)} = 0$).}
When the coefficient of the highest-derivative term vanishes, the $\Box^2$ contribution to $c(\Box)$ disappears, reducing the differential order of the governing equation for $R^{(1)}$ from fourth to second order. Correspondingly, one of the scalar degrees of freedom
may cease to be an independent propagating mode. In this limit, the mass expressions in Eq.~\eqref{masas-campos}, which contain the vanishing combination in their denominators, become singular, and the degenerate theory must instead be analyzed using the reduced-order field equations.

\paragraph{Both Scalar Modes Nondynamical ($c(\Box) = a$ and $e(\Box) = 0$).}
A further degenerate regime occurs when $c(\Box) = a$ and $e(\Box) = 0$, corresponding to
\[
2\mathcal{A} f_{,\mathcal{R}}^{(0)}
+ f_{,RR}^{(0)} - \mathcal{B} f_{,R}^{(0)} = 0,
\qquad
\mathcal{A}^2 f_{,\mathcal{R}}^{(0)}
- \mathcal{B} f_{,RR}^{(0)} = 0.
\]
Under these conditions, neither scalar mode describes an independent propagating degree of freedom.

\subsection{Absence of ghosts and tachyons
in \texorpdfstring
{$f(R,\mathcal{R})$}~ theories
}
\label{constraints}
In this subsection, the conditions under which the
\(f(R,\mathcal{R})\) theories are free from ghost and tachyonic instabilities
are derived. Starting from the linearized field equations \eqref{expanded-field-eq-FO}, we invert the corresponding kinetic operator to obtain the propagator associated with the graviton and the additional modes arising in the theory. The linearized field equations 
\eqref{expanded-field-eq-FO}
can be written as follows 
\begin{equation}
\mathcal{O}_{\mu\nu\lambda\sigma}h^{\lambda\sigma}=2\kappa_{\mathrm{eff}}\, T^{\mathrm{eff}}_{\mu\nu}.
\end{equation}
Following the formalism presented in \cite{Biswas2011,Koivisto2013}, 
the linearized field equations are written in momentum space, and we invert them using the Barnes--Rivers
spin-projector basis. We define the transverse and longitudinal operators
\begin{equation}
    \theta_{\mu\nu}=\eta_{\mu\nu}-\frac{k_\mu k_\nu}{k^2},
    \qquad
    \omega_{\mu\nu}=\frac{k_\mu k_\nu}{k^2},
\end{equation}
from which the spin-2 and scalar spin-0 projectors are given by
\begin{equation}
    \mathcal{P}^2_{\mu\nu\lambda\sigma}
    =
    \frac{1}{2}
    \left(
        \theta_{\mu\lambda}\theta_{\nu\sigma}
        +
        \theta_{\mu\sigma}\theta_{\nu\lambda}
    \right)
    -
    \frac{1}{3}\theta_{\mu\nu}\theta_{\lambda\sigma},
\end{equation}
and
\begin{equation}
    \mathcal{P}^0_{s\,\mu\nu\lambda\sigma}
    =
    \frac{1}{3}\theta_{\mu\nu}\theta_{\lambda\sigma}.
\end{equation}
The remaining projectors, associated with vector and longitudinal scalar
components, do not contribute and are therefore omitted here. In this representation, one finds
\begin{eqnarray}
    \nonumber h_{\mu\nu}&=&2\kappa_{\mathrm{eff}}\, \mathcal{O}^{-1\,\lambda\sigma}_{\mu\nu}T^{\mathrm{eff}}_{\lambda\sigma}\nonumber\\&=&2\kappa_{\mathrm{eff}}\left[\frac{\mathcal{P}^{2~~\lambda\sigma}_{\mu\nu}}{k^2}+
    \frac{a}
    {k^2\left(a-3c(-k^2)\right)}
    \mathcal{P}^{0~~~\lambda\sigma}_{s\,\mu\nu}\right]T^{\mathrm{eff}}_{\lambda\sigma}\\&=&2\kappa_{\mathrm{eff}}\left[\frac{\mathcal{P}^2_{\mu\nu\lambda\sigma}}{k^2}
    +
    \frac{a\,(1-3\mathcal{B}k^2)}
    {k^2\left(a-3c(-k^2)\right)}
    \mathcal{P}^0_{s\,\mu\nu\lambda\sigma}\right]T^{(1)\,\lambda\sigma}.
\end{eqnarray}
Then, the propagator is given by\footnote{For notational simplicity, the tensor indices of the propagator
have been suppressed.}
\begin{equation}
    \Pi_{f(R,\mathcal{R})}
    =
    \frac{\mathcal{P}^2}{k^2}
    +
    \frac{a\,(1-3\mathcal{B}k^2)}
    {k^2\left(a-3c(-k^2)\right)}
    \mathcal{P}^0_s.
\end{equation}
Using the standard $\rm GR$ propagator,
 $$   \Pi_{\rm GR}
    =
    \frac{\mathcal{P}^2}{k^2}
    -
    \frac{1}{2}\frac{\mathcal{P}^0_s}{k^2},
$$
this expression may also be written as
\begin{equation}
    \Pi_{f(R,\mathcal{R})}=\Pi_{GR}+\frac{3}{2}\frac{\left(a-c(-k^2)-2\mathcal{B}ak^2\right)}{k^2\left(a-3c(-k^2)\right)}\mathcal{P}^0_s,
\end{equation}
or, equivalently, in terms of the scalar masses,
\begin{equation}
    \Pi_{f(R,\mathcal{R})}=\Pi_{GR}+\frac{3}{4}\frac{m_+^2 m_-^2\left(a-c(-k^2)-2\mathcal{B}ak^2\right)}{a\,k^2\,\left(m_+^2-m_-^2\right)\left(k^2+m_+^2\right)}\mathcal{P}^0_s-\frac{3}{4}\frac{m_+^2 m_-^2\left(a-c(-k^2)-2\mathcal{B}ak^2\right)}{a\,k^2\,\left(m_+^2-m_-^2\right)\left(k^2+m_-^2\right)}\mathcal{P}^0_s.
    \label{propagator}
\end{equation}
This equation agrees with the results presented in the previous section:\(f(R,\mathcal{R})\) theories contain an additional spin-0 sector, which has two poles, each corresponding to an extra scalar degree of freedom.

Since \(c(m_\pm^2)=a/3\), the residues at the poles are
\begin{equation}
    z_+=-\frac{1}{2}\frac{m_-^2\left(1+3\mathcal{B}m_+^2\right)}{\left(m_+^2-m_-^2\right)},~\text{for}~k^2=-m_+^2,
\end{equation}
and
\begin{equation}
    z_-=\frac{1}{2}\frac{m_+^2\left(1+3\mathcal{B}m_-^2\right)}{\left(m_+^2-m_-^2\right)},~\text{for}~k^2=-m_-^2.
\end{equation}

To ensure that the propagating degrees of freedom are free from both tachyonic instabilities and ghostlike behavior, the following inequalities
\cite{Koivisto2013,Green2012}
\begin{equation}
m_{\pm}^{2}>0 \quad \text{and} \quad z_{\pm}>0,
\end{equation}
must be satisfied. We also impose the condition
\begin{equation}
    f_{,R}^{(0)}+f_{,\mathcal{R}}^{(0)}>0,
    \label{a-condition}
\end{equation}
so that the effective gravitational coupling 
 $$   \kappa_{\mathrm{eff}}=\frac{8\pi G}{f_{,R}^{(0)}+f_{,\mathcal{R}}^{(0)}}
$$
is positive,
ensuring that ordinary matter sources generate an attractive gravitational interaction in the weak-field limit.

Let us start with the conditions \(m_\pm^2>0\). From the product of the squared masses, we obtain
\begin{equation}
    m_+^2 m_-^2=\frac{\left(f_{,R}^{(0)}+f_{,\mathcal{R}}^{(0)}\right)}{9\left(\mathcal{A}^2\,f_{,\mathcal{R}}^{(0)}
- \mathcal{B}\,f_{,RR}^{(0)}\right)}.
\label{masses-product}
\end{equation}
Since Eq.~\eqref{a-condition} implies that the numerator of Eq.\eqref{masses-product} is strictly positive and the denominator of the latter is non-vanishing for a physically viable theory, the product $m_+^2 m_-^2$ is non-zero, implying that neither mass can vanish under these assumptions.
Then, positivity of the product of the masses requires that
\begin{equation}
    \left(\mathcal{A}^2\,f_{,\mathcal{R}}^{(0)}
- \mathcal{B}\,f_{,RR}^{(0)}\right)>0.
\end{equation}
Using the definitions of \(\mathcal{A}\) and \(\mathcal{B}\), this condition can be rewritten as
\begin{equation}
    \frac{1}{f_{,\mathcal{R}}^{(0)}}\left[\left(f_{,R\mathcal{R}}^{(0)}\right)^2
- f^{(0)}_{,\mathcal{R}\mathcal{R}}\,f_{,RR}^{(0)}\right]>0.
\label{59}
\end{equation}
From the sum of the masses, we have
\begin{equation}
    m_+^2+m_-^2=
    \frac{2\mathcal{A}\,f_{,\mathcal{R}}^{(0)}+f_{,RR}^{(0)}-\mathcal{B}\,f_{,R}^{(0)}}
    {3\left(\mathcal{A}^2\,f_{,\mathcal{R}}^{(0)}
- \mathcal{B}\,f_{,RR}^{(0)}\right)}.
\end{equation}
Since the denominator is already positive, positivity of the sum implies
\begin{equation}
    2\mathcal{A}\,f_{,\mathcal{R}}^{(0)}+f_{,RR}^{(0)}-\mathcal{B}\,f_{,R}^{(0)}>0.
    \label{62}
\end{equation}
Moreover, the difference of the masses is
\begin{equation}
    m_+^2-m_-^2=\frac{\sqrt{ \left(\mathcal{B}\,f_{,R}^{(0)} - 2\mathcal{A}\,f_{,\mathcal{R}}^{(0)} - f_{,RR}^{(0)}\right)^2-4\left(f_{,\mathcal{R}}^{(0)} + f_{,R}^{(0)}\right)\left(\mathcal{A}^2\,f_{,\mathcal{R}}^{(0)}
- \mathcal{B}\,f_{,RR}^{(0)}\right)}}{3\left(\mathcal{A}^2\,f_{,\mathcal{R}}^{(0)}
- \mathcal{B}\,f_{,RR}^{(0)}\right)}>0,
\end{equation}
so that \(m_+^2>m_-^2\).

We now turn to the residue conditions. Since \(m_+^2>m_-^2\), positivity of \(z_\pm\) is equivalent to
\begin{equation}
-\left(1+3\mathcal{B}m_+^2\right)>0,
\qquad
\left(1+3\mathcal{B}m_-^2\right)>0.
\label{residues-inequalities}
\end{equation}
Adding these two inequalities gives
\begin{equation}
    3\mathcal{B}\left(m_-^2-m_+^2\right)>0,
\end{equation}
and therefore
\begin{equation}
    \mathcal{B}<0.
    \label{64}
\end{equation}

The product of the inequalities in \eqref{residues-inequalities} yields
\begin{equation}
    -\left(1+3\mathcal{B}m_+^2\right)\left(1+3\mathcal{B}m_-^2\right)>0,
\end{equation}
which can be rewritten as
\begin{equation}
    -\frac{\left(f_{,R\mathcal{R}}^{(0)}+f^{(0)}_{,\mathcal{R}\mathcal{R}}\right)^2}{\left[\left(f_{,R\mathcal{R}}^{(0)}\right)^2
- f^{(0)}_{,\mathcal{R}\mathcal{R}}\,f_{,RR}^{(0)}\right]}>0.
\end{equation}
Hence,
\begin{equation}
    \left(f_{,R\mathcal{R}}^{(0)}\right)^2
- f^{(0)}_{,\mathcal{R}\mathcal{R}}\,f_{,RR}^{(0)}<0,
\label{65}
\end{equation}
and
\begin{equation}
    f_{,R\mathcal{R}}^{(0)}+f^{(0)}_{,\mathcal{R}\mathcal{R}}\neq0.
    \label{70}
\end{equation}
Combining \eqref{65} with \eqref{59}, we conclude that
\begin{equation}
    f_{,\mathcal{R}}^{(0)}<0.
\end{equation}
Since \(\mathcal{B}=f^{(0)}_{,\mathcal{R}\mathcal{R}}/f_{,\mathcal{R}}^{(0)}<0\) and \(f_{,\mathcal{R}}^{(0)}<0\), it follows that
\begin{equation}
    f^{(0)}_{,\mathcal{R}\mathcal{R}}>0.
\end{equation}
Finally, Eq.~\eqref{65} leads to
\begin{equation}
    f^{(0)}_{,RR}>0.
\end{equation}

The conditions required to avoid both tachyonic and ghostlike instabilities 
can therefore be summarized as
\begin{equation}
f^{(0)}_{,RR}>0,
\qquad
f^{(0)}_{,\mathcal{R}\mathcal{R}}>0,
\qquad
f^{(0)}_{,\mathcal{R}}<0,
\qquad
\left(f_{,R\mathcal{R}}^{(0)}\right)^2<f^{(0)}_{,\mathcal{R}\mathcal{R}}\,f^{(0)}_{,RR},
\end{equation}
supplemented by
\begin{equation}
    f_{,R\mathcal{R}}^{(0)}+f^{(0)}_{,\mathcal{R}\mathcal{R}}\neq0,
    \qquad
    f_{,R}^{(0)}+f_{,\mathcal{R}}^{(0)}>0.
\end{equation}
These conditions coincide with those derived in \cite{Bombacigno2019}, except for the additional condition in Eq.\eqref{70}. Our results also show that the condition \(f_{,\mathcal{R}\mathcal{R}}\neq 0\)
is of fundamental importance in the avoidance of ghost and tachyonic instabilities\footnote{These results were also derived as a particular case of extended hybrid metric-Palatini theories, which incorporate Ricci-squared invariants \cite{RamirezMelgarejo2026}.}.

Under the conditions derived above, Eq.~\eqref{62} is automatically satisfied. Moreover, the parameter \(\mathcal{B}\) is constrained to lie in the interval
\begin{equation}
    -\frac{1}{3m_-^2}<\mathcal{B}<-\frac{1}{3m_+^2}.
\end{equation}

\section{Weak-field phenomenology} \label{section 5}
We now turn to the weak-field phenomenology of GHT. In
the following, the modified Newtonian potential generated by a static source and the corresponding post-Newtonian parameter \(\gamma\)
will be derived. These quantities provide a direct link between the linearized theory and observational tests in the Solar System.

\subsection{Gravitational potential and the post-Newtonian parameter \texorpdfstring{$\gamma$}~} \label{subsection 5.1}

To compute the gravitational potential, it is convenient to introduce the functions 
$\mathcal{H}^{(m_i)}_{\mu\nu}(\bm{r},t)$, defined by \cite{Bueno2016} 
\begin{equation}
    \left(\Box-m_i^2\right)\mathcal{H}^{(m_i)}_{\mu\nu}(\bm{r},t)=-4\pi T_{\mu\nu}(\bm{r},t),
    \label{General-KG}
\end{equation}
where $m_i=m_+,m_-,0$.
It then follows from Eqs. \eqref{eq:mov-phi}, \eqref{eq:mov-psi}, and \eqref{eq:mov-hat-h} that\footnote{Since \(\hat{h}_{\mu\nu}\) is transverse, \(\mathcal{H}^{(0)}_{\mu\nu}\) is also transverse.}
\begin{eqnarray}\nonumber
    \hat{h}_{\mu\nu}&=&\frac{\kappa_{\mathrm{eff}}}{2\pi}\left[\mathcal{H}^{(0)}_{\mu\nu}-\mathcal{B}\,\partial_\mu\partial_\nu\mathcal{H}^{(0)}+\mathcal{B}\,\eta_{\mu\nu}\Box\mathcal{H}^{(0)}\right],\\ \phi&=&\frac{\kappa_{\mathrm{eff}}}{2\pi}\left[\mathcal{H}^{(m_+)}+3\mathcal{B}\,\Box\mathcal{H}^{(m_+)}\right],\\ \nonumber \psi&=&\frac{\kappa_{\mathrm{eff}}}{2\pi}\left[\mathcal{H}^{(m_-)}+3\mathcal{B}\,\Box\mathcal{H}^{(m_-)}\right].
    \label{mov-hath-gamma-varphi}
\end{eqnarray}
Substituting these expressions into the metric decomposition \eqref{h-grados-de-libertad}, and performing a gauge transformation of the form
%
$$
    h^N_{\mu\nu}\equiv h_{\mu\nu}-\partial_{(\mu}\xi_{\nu)},
$$
where the superscript \(N\) denotes the Newtonian gauge, with
\begin{equation}
    \xi_{\nu}\equiv\frac{\kappa_{\text{eff}}}{2\pi}\,\partial_{\nu}\left[\frac{m_+^2\left(1+3\mathcal{B}m_-^2\right)}{3 m_-^2 \left(m_+^2-m_-^2\right)}\mathcal{H}^{(m_-)}-\frac{m_-^2\left(1+3\mathcal{B}m_+^2\right)}{3 m_+^2 \left(m_+^2-m_-^2\right)}\mathcal{H}^{(m_+)}-\left(\frac{\left(m_+^2+m_-^2\right)}{3m_+^2 m_-^2}+\mathcal{B}\right)\mathcal{H}^{(0)}\right],
\end{equation}
it follows that
\begin{equation}
      h^N_{\mu\nu} = \frac{\kappa_{\text{eff}}}{4\pi}\left( 2\mathcal{H}^{(0)}_{\mu\nu}-\eta_{\mu\nu}\mathcal{H}^{(0)}+\frac{1}{3}\frac{\eta_{\mu\nu}}{ m^{2}_{+}-m^{2}_{-}}\left[m^{2}_{+}\left(1+3\mathcal{B} m_{-}^{2}\right)\mathcal{H}^{(m_-)}- m^{2}_{-}\left(1+3\mathcal{B} m_{+}^{2}\right)\mathcal{H}^{(m_{+})}\right]\right).
      \label{Newtonian-h-grados-de-libertad}
\end{equation}
Restricting to static configurations, Eq.~\eqref{General-KG} becomes
\begin{equation}
    \left(\nabla^2-m_i^2\right)\mathcal{H}^{(m_i)}_{\mu\nu}\left(\bm{r}\right)=-4\pi T_{\mu\nu}(\bm{r}),
\end{equation}
which is an inhomogeneous Helmholtz equation. Its solution is given by \cite{Bueno2016}
\begin{equation}
\label{solhel}
    \mathcal{H}^{(m_i)}_{\mu\nu}(\bm{r})=\int d^3r^{\prime} ~\frac{T_{\mu\nu}(\bm{r}^{\prime})}{|\bm{r}-\bm{r}^{\prime}|}e^{-m_i|\bm{r}-\bm{r}^{\prime}|}.
\end{equation}
%
%
Let us apply these expressions 
to a source with radius $R_*$, described in the Newtonian regime by 
\begin{equation}
   T_{00}= \varrho(r)=\rho(r)\Theta(R_\star-r).
\end{equation}
Standard calculations lead to 
\begin{equation}
    \mathcal H^{(m_i)}_{00}(r)
    =
    \mathcal{F}(m_i R_{\star})\:\frac{M}{r}e^{-m_ir}\,
    ,
    \qquad r>R_\star,
    \label{H00-exterior-general-density}
\end{equation}
where the density-dependent form factor is
\begin{equation}
    \mathcal{F}(m_i R_{\star})
    =
    \frac{4\pi}{M}
    \int_0^{R_\star}dr^{\prime}\,r^{\prime2}\rho(r^{\prime})
    \frac{\sinh(m_ir^{\prime})}{m_ir^{\prime}}.
    \label{general-density-form-factor}
\end{equation}
Here $M$ denotes the mass integral of the source entering the linearized solution. It is not identified a priori with the gravitational mass inferred from orbital motion

With the convention
 $$   \mathcal{H}^{(m_i)}(r)=-\mathcal{H}^{(m_i)}_{00}(r),
$$
it follows that
\begin{equation}
    \mathcal{H}^{(m_i)}(r)
    =
    -\mathcal{F}(m_i R_{\star})\frac{M}{r}e^{-m_ir}.
    \label{H-exterior-general-density}
\end{equation}
Using the series expansion for $m_i R_{\star}\ll1$ 
of the factor multiplying $r'^2\:\rho(r')$ in Eq.
\eqref{general-density-form-factor}, 
we see that
\begin{equation}
     \nonumber\mathcal{F}(m_i R_{\star})\simeq1
\end{equation}
for $m_i R_{\star}\ll1$.

In the case of a homogeneous sphere of radius \(R_\star\) and total mass \(M\), 
$$
    T_{00}=\rho(r)=\rho_0 \Theta(R_\star-r), ~~~~  {\rm where}~~~~\rho_0=\frac{3M}{4\pi R_\star^3},
$$
it follows that, for \(r>R_\star\),
\begin{equation}
    \mathcal{H}^{(m_i)}(r)=-\mathcal{H}^{(m_i)}_{00}(r)=-\mathcal{F}(m_i R_\star)\frac{M}{r}e^{-m_i r},
\end{equation}
where
\begin{equation}
   \mathcal{F}(m_i R_\star)=\frac{3}{(m_i R_\star)^3}[m_i R_\star\cosh{(m_i R_\star)}-\sinh{(m_i R_\star)}].
\end{equation}
The function \(\mathcal{F}(m_i R_\star)\) behaves as \(\mathcal{F}(m_i R_\star)\approx \frac{3}{2}\frac{1}{(m_i R_\star)^2}e^{m_i R_\star}\) for \(m_i R_\star \gg 1\), while \(\mathcal{F}(m_i R_\star)\to 1\) for \(m_i R_\star\ll 1\).

Substituting these results into Eq.~\eqref{Newtonian-h-grados-de-libertad}, we obtain
\begin{equation}
    h^N_{00} =  \frac{2MG}{ar}\left[ 1+ \frac{1}{3}\frac{m^{2}_{+}}{ m^{2}_{+}-m^{2}_{-}}\left(1+3\mathcal{B} m_-^2\right)\mathcal{F}(m_{-} R_\star)e^{-m_{-} r}- \frac{1}{3}\frac{ m^{2}_{-}}{m^{2}_{+}-m^{2}_{-}}\left(1+3\mathcal{B} m_{+}^{2}\right)\mathcal{F}(m_{+}R_\star)e^{-m_{+}r}\right],
    \label{h^N_00}
\end{equation}
and
\begin{equation}
h^N_{ij}=
    \frac{2MG}{ar}\delta_{ij}\left[1- \frac{1}{3}\frac{ m^{2}_{+}}{ m^{2}_{+}-m^{2}_{-}}\left(1+3\mathcal{B}m^{2}_{-}\right)\mathcal{F}(m_{-} R_\star)e^{-m_{-} r}+ \frac{1}{3}\frac{ m^{2}_{-}}{ m^{2}_{+}-m^{2}_{-}}(1+3\mathcal{B}m^{2}_{+})\mathcal{F}(m_{+}R_\star)e^{-m_{+}r}\right].
    \label{h^N_ij}
\end{equation}
Using \(g^{N}_{\mu\nu}=\eta_{\mu\nu}+h^{N}_{\mu\nu}\), the line element can be written as
\begin{equation}
    ds^2=-(1+2U(r))dt^2+(1-2\gamma(r)U(r))\delta_{ij}dx^i dx^j,
\label{newtonian-ds}    
\end{equation}
where the generalized Newtonian potential \(U(r)\) and the 
so-called 
post-Newtonian parameter \(\gamma(r)\) are given by
\begin{equation}
    U(r)=-\frac{G M}{ar} \left[ 1+ \frac{1}{3}\frac{m^{2}_{+}}{ m^{2}_{+}-m^{2}_{-}}\left(1+3\mathcal{B} m_-^2\right)\mathcal{F}(m_{-} R_\star)e^{-m_{-} r}- \frac{1}{3}\frac{ m^{2}_{-}}{m^{2}_{+}-m^{2}_{-}}\left(1+3\mathcal{B} m_{+}^{2}\right)\mathcal{F}(m_{+}R_\star)e^{-m_{+}r}\right],
    \label{potential}
\end{equation}
and
\begin{equation}
\gamma(r)
=
\frac{
1
-\frac{1}{3}\frac{m_{+}^{2}}{m_{+}^{2}-m_{-}^{2}}\left(1+3\mathcal{B}m_{-}^{2}\right)\mathcal{F}(m_{-}R_\star)e^{-m_{-}r}
+\frac{1}{3}\frac{m_{-}^{2}}{m_{+}^{2}-m_{-}^{2}}\left(1+3\mathcal{B}m_{+}^{2}\right)\mathcal{F}(m_{+}R_\star)e^{-m_{+}r}
}{
1
+\frac{1}{3}\frac{m_{+}^{2}}{m_{+}^{2}-m_{-}^{2}}\left(1+3\mathcal{B}m_{-}^{2}\right)\mathcal{F}(m_{-}R_\star)e^{-m_{-}r}
-\frac{1}{3}\frac{m_{-}^{2}}{m_{+}^{2}-m_{-}^{2}}\left(1+3\mathcal{B}m_{+}^{2}\right)\mathcal{F}(m_{+}R_\star)e^{-m_{+}r}
}.
\label{parameter-gamma}
\end{equation}
Notice that the potential in Eq.~\eqref{potential} is invariant under the exchange \(m_+\leftrightarrow m_-\), as expected from the symmetry of the factorized scalar sector.

For later use, it is convenient to define the Yukawa amplitudes
\[
\alpha_-\equiv\frac{1}{3}\frac{m_+^2}{m_+^2-m_-^2}(1+3\mathcal B m_-^2),
\qquad
\alpha_+\equiv-\frac{1}{3}\frac{m_-^2}{m_+^2-m_-^2}(1+3\mathcal B m_+^2),
\]
which are related to the scalar residues in the following way:
 $$   \alpha_-\equiv \frac{2}{3}z_-,
    \qquad
    \alpha_+\equiv \frac{2}{3}z_+.
$$
Hence, the generalized Newtonian potential and the parameter \(\gamma\) can be written as
\begin{equation}
    U(r)=-\frac{G M}{ar} \left[ 1+ \alpha_-\,\mathcal{F}(m_{-} R_\star)e^{-m_{-} r}+ \alpha_+\,\mathcal{F}(m_{+}R_\star)e^{-m_{+}r}\right],
    \label{potential-2}
\end{equation}
and
\begin{equation}
\gamma(r)
=
\frac{
1
-\alpha_-\,\mathcal{F}(m_{-}R_\star)e^{-m_{-}r}
-\alpha_+\,\mathcal{F}(m_{+}R_\star)e^{-m_{+}r}
}{
1
+\alpha_-\,\mathcal{F}(m_{-}R_\star)e^{-m_{-}r}
+\alpha_+\,\mathcal{F}(m_{+}R_\star)e^{-m_{+}r}
}.
\label{parameter-gamma-2}
\end{equation}

To link these expressions with observations, we define the dimensionless scalar contribution
\begin{equation}
\Delta(r)\equiv
\alpha_-\,\mathcal F(m_-R_\star)e^{-m_-r}
+\alpha_+\,\mathcal F(m_+R_\star)e^{-m_+r}.
\label{scalar-contribution}
\end{equation}
The exterior potential in Eq.~\eqref{potential-2} can then be written as
\begin{equation}
U(r)=-\frac{GM}{ar}[1+\Delta(r)].
\label{potential-delta}
\end{equation}
At Newtonian order, the acceleration of a slowly moving test body follows from $-\boldsymbol{\nabla}U$:
\begin{equation}
\ddot{\boldsymbol r}
=
-\frac{GM}{ar^2}
\left[1+\Delta(r)-r\Delta'(r)\right]
\hat{\boldsymbol r}.
\label{orbital-acceleration}
\end{equation}
If the scalar contribution is approximately constant over the relevant observational region, $\Delta(r)\simeq\Delta_0$, and $r\Delta'(r)$ is negligible at the required accuracy, the acceleration reduces to an inverse-square law. The gravitational parameter inferred from the leading orbital dynamics is then
\begin{equation}
\mu_{\rm obs}\equiv\frac{GM}{a}(1+\Delta_0).
\label{constant-scalar-calibration}
\end{equation}
Consequently, the potential becomes
\begin{equation}
U(r)\simeq-\frac{\mu_{\rm obs}}{r}.
\label{calibrated-newtonian-potential}
\end{equation}
Thus, an approximately constant scalar contribution can be absorbed into the gravitational parameter inferred from orbital motion and cannot be constrained independently through the Newtonian orbital normalization alone.

Using the same normalization, the linearized exterior metric in Eq.~\eqref{newtonian-ds} takes the form
\begin{equation}
ds^2\simeq
-\left(1-\frac{2\mu_{\rm obs}}{r}\right)dt^2
+\left(1+2\gamma_{\rm obs}
\frac{\mu_{\rm obs}}{r}\right)\delta_{ij}dx^i dx^j,
\label{calibrated-linear-metric}
\end{equation}
where
\begin{equation}
\gamma_{\rm obs}
=
\frac{1-\Delta_0}{1+\Delta_0}.
\label{gamma-observed}
\end{equation}

In this approximately constant-scalar regime, $\gamma_{\rm obs}$ has the usual linear PPN interpretation when the same normalization applies over the region relevant for light propagation. Although the leading orbital potential has the Newtonian form after calibration, the spatial metric may still differ from its general-relativistic form.

If $\Delta(r)$ varies appreciably over the relevant region, its radial dependence cannot in general be absorbed into a single constant gravitational parameter.

The combinations
$\alpha_i\mathcal F(m_iR_\star)e^{-m_ir}$
appearing in Eqs.~\eqref{potential-2} and
\eqref{parameter-gamma-2} measure the scalar contributions to
the metric potentials relative to the coefficient $GM/a$.
Their suppression provides a sufficient condition for
recovering the metric potentials without appreciable scalar
contributions. However, such suppression is not necessary for
the leading Newtonian orbital potential after calibration,
as shown by Eq.~\eqref{calibrated-newtonian-potential}. 
The factor
\(\mathcal F(m_iR_\star)\) encodes the details of the source,
whereas \(e^{-m_ir}\) gives the Yukawa decay outside it. 
In the following, we will use the condition that the contribution associated with the scalar mode \(m_i\) is negligible whenever
\begin{equation}
   {\alpha_i} \mathcal F(m_iR_\star)e^{-m_ir}\ll1.
    \label{mode-suppression-condition}
\end{equation}

In the point-like limit, \(\mathcal{F}(m_i R_\star)=1\), and Eqs.~\eqref{potential-2} and \eqref{parameter-gamma-2} reduce to
\begin{equation}
    U(r)=-\frac{G M}{ar} \left[ 1+ \alpha_-\,e^{-m_{-} r}+ \alpha_+\,e^{-m_{+}r}\right],
    \label{newton-potential}
\end{equation}
and
\begin{equation}
\gamma(r)=\frac{
1
-\alpha_-\,e^{-m_{-}r}
-\alpha_+\,e^{-m_{+}r}
}{
1
+\alpha_-\,e^{-m_{-}r}
+\alpha_+\,e^{-m_{+}r}
}.
\label{ppn-gamma}
\end{equation}
Hence, in this limit the usual Newtonian potential is modified by two Yukawa-like corrections associated with the two additional scalar modes, whose masses are determined by Eq.~\eqref{masas-campos}.
Each correction amplitude is proportional to the corresponding scalar residue. These results are in line with the ones obtained in \cite{Bombacigno2019,Rosa2021},

The results for the point-like source 
in the case of the family of hybrid theories defined by 
\cite{Capo2015}
\[
f(R,\mathcal{R})=R+\xi(\mathcal{R}),
\]
are recovered by taking the limits \(m_+\rightarrow\infty\) and \(m_-R_\star\ll1\) in Eqs.~\eqref{potential} and \eqref{parameter-gamma}. In addition, for this class of models the scalar mass is given by
\[
m_-^2=-\frac{1+\xi^{(0)}_{,\mathcal{R}}}{3\mathcal{B}},
\]
which follows from Eq.~\eqref{eq:O-in-GR} after setting \(f(R,\mathcal{R})=R+\xi(\mathcal{R})\), taking the trace, and factorizing the resulting operator. The resulting expressions are
\begin{equation}
    U(r)=-\frac{G M}{\left(1+\xi^{(0)}_{,\mathcal{R}}\right)r} \left[ 1- \frac{\xi^{(0)}_{,\mathcal{R}}}{3}e^{-m_{-} r}\right],
    \label{potential-fX}
\end{equation}
and
\begin{equation}
\gamma(r)
=
\frac{
1
+\frac{\xi^{(0)}_{,\mathcal{R}}}{3}e^{-m_{-}r}}{
1
-\frac{\xi^{(0)}_{,\mathcal{R}}}{3}e^{-m_{-}r}},
\label{parameter-gamma-fX}
\end{equation}
which coincide with those presented in \cite{Harko2011}.

The ratio $\gamma(r)$ in Eq.~\eqref{parameter-gamma-2} is generally position-dependent and cannot automatically be identified with the constant parameter of the standard PPN formalism~\cite{Will2014}. In the approximately constant-scalar regime, the orbital calibration in Eq.~\eqref{constant-scalar-calibration} gives the metric ratio $\gamma_{\rm obs}$ in Eq.~\eqref{gamma-observed}. When the scalar contribution varies appreciably on short scales, light-propagation predictions must instead be obtained from the radially dependent metric~\cite{Toniato2021,Huang2024}. Appendix~\ref{appendix} discusses
the normalization of the combination entering the null-geodesic equation.

\subsection{Weak-field limit in different 
regimes for the 
scalar masses}\label{sec:mass-hierarchy}

Let us now analyze the weak-field potential 
assuming different relations between the two
scalar masses \(m_+\) and \(m_-\). Starting from Eq.~\eqref{potential-2}, we
recall that, in the healthy sector, the absence of tachyonic and ghostlike
instabilities requires
\begin{equation}
    -\frac{1}{3m_-^2}<\mathcal B<-\frac{1}{3m_+^2},
    \qquad
    m_+>m_-.
    \label{B-interval}
\end{equation}
The former inequality  implies that both Yukawa amplitudes are positive, namely
\(\alpha_->0\) and \(\alpha_+>0\).

It is convenient to parametrize the position of \(\mathcal B\) inside the
allowed interval by defining
\begin{equation}
    \chi\equiv
    \frac{1+3\mathcal Bm_-^2}{1-m_-^2/m_+^2}.
    \label{chi-def}
\end{equation}
The bounds \eqref{B-interval} imply that $0<\chi<1$. Equivalently,
\begin{equation}
    \mathcal B
    =
    -\frac{1}{3m_-^2}
    +
    \chi
    \left(
    \frac{1}{3m_-^2}
    -
    \frac{1}{3m_+^2}
    \right).
    \label{B-chi}
\end{equation}
Thus, the limiting values \(\chi\to0\) and \(\chi\to1\) correspond respectively
to the lower and upper endpoints of the allowed interval for \(\mathcal B\).

In terms of \(\chi\), the Yukawa amplitudes take the simple form
\begin{equation}
    \alpha_-=\frac{\chi}{3},
    \qquad
    \alpha_+=\frac{1-\chi}{3}.
    \label{alpha-chi}
\end{equation}
Hence
\begin{equation}
    \alpha_-+\alpha_+=\frac13.
    \label{alpha-sum}
\end{equation}
Notice that the parameter \(\mathcal B\) distributes the total scalar amplitude \(1/3\) between the scalar modes. The consequences of this distribution depend crucially on the relation between the masses, as will be
discussed next.

\subsubsection{Nearly degenerate regime}

First, we consider the case in which the two scalar masses are almost equal.
We write
\begin{equation}
    m_+^2=m_0^2+\delta,
    \qquad
    m_-^2=m_0^2-\delta,
    \qquad
    0<\delta\ll m_0^2.
    \label{degenerate-masses}
\end{equation}
Then
\begin{equation}
    m_+^2-m_-^2=2\delta,
\end{equation}
and the stability interval in Eq.\eqref{B-interval} becomes
\begin{equation}
    -\frac{1}{3(m_0^2-\delta)}
    <
    \mathcal B
    <
    -\frac{1}{3(m_0^2+\delta)}.
    \label{B-degenerate-exact}
\end{equation}
Expanding for \(\delta\ll m_0^2\), one obtains
\begin{equation}
    -\frac{1}{3m_0^2}
    -\frac{\delta}{3m_0^4}
    +\mathcal{O}\!\left(\frac{\delta^2}{m_0^6}\right)
    <
    \mathcal B
    <
    -\frac{1}{3m_0^2}
    +\frac{\delta}{3m_0^4}
    +\mathcal{O}\!\left(\frac{\delta^2}{m_0^6}\right).
    \label{B-degenerate-expanded}
\end{equation}
Thus, in the nearly degenerate regime, \(\mathcal B\) is restricted to a
narrow interval around $
    \mathcal B_0=-\frac{1}{3m_0^2}$. Indeed, using Eq.~\eqref{B-chi}, one finds
\begin{equation}
    \mathcal B
    =
    -\frac{1}{3m_0^2}
    +
    \frac{2\chi-1}{3}\frac{\delta}{m_0^4}
    +\mathcal{O}\!\left(\frac{\delta^2}{m_0^6}\right).
    \label{B-degenerate-chi}
\end{equation}
The endpoints \(\chi=0\) and \(\chi=1\) reproduce the two endpoints of
Eq.~\eqref{B-degenerate-expanded}.

Since
\begin{equation}
    m_\pm
    =
    m_0
    \pm
    \frac{\delta}{2m_0}
    +
    \mathcal{O}\!\left(\frac{\delta^2}{m_0^3}\right),
\end{equation}
the Yukawa factors 
can be expanded 
around \(m_0\). Using Eq.~\eqref{alpha-chi},
it follows that
\begin{equation}
    U(r)\simeq
    -\frac{GM}{ar}
    \left[
    1+\frac13\mathcal F(m_0 R_\star)e^{-m_0r}+\frac{(1-2\chi)\,\delta}{6m_0}\frac{\partial}{\partial m_0}\left[\mathcal F(m_0 R_\star)e^{-m_0r}\right]+\mathcal{O}\!\left[\left(\frac{\delta}{m_0^2}\right)^2\right]
    \right].
\end{equation}
Therefore, to leading order in the small parameter \(\delta/m_0^2\), the
dependence on \(\chi\) drops out and the potential reduces to
\begin{equation}
    U(r)\simeq
    -\frac{GM}{ar}
    \left[
    1+\frac13\mathcal F(m_0 R_\star)e^{-m_0r}
    \right],
    \qquad
    m_+^2\simeq m_-^2\simeq m_0^2.
    \label{potential-degenerate}
\end{equation}
Hence, in the nearly degenerate regime, the two scalar modes behave, at leading
order, as a single Yukawa correction with total amplitude \(1/3\).

Let us now consider the long-range limit, in which the common scalar Compton
wavelength is larger than every length scale relevant in the region 
outside the source. If \(\ell_{\rm max}\) denotes the largest distance at which the
potential is probed, with \(\ell_{\rm max}>R_\star\), this regime is
characterized by 
\(m_0 R_{\star}\ll1\), 
\(m_0\ell_{\rm max}\ll1\).
Therefore
\(\mathcal F(m_0R_\star)\simeq1\) and \(e^{-m_0r}\simeq1\).
Hence,
$$
    \mathcal F(m_0R_\star)e^{-m_0r}\simeq1 
$$
for $R_\star \leq r \leq \ell_{\rm max}$.
Equation~\eqref{potential-degenerate} then gives
\begin{equation}
    U(r)\simeq
    -\frac{4GM}{3ar}.
    \label{potential-degenerate-four-thirds}
\end{equation}
The factor $4/3a$ represents an enhancement relative to the bare coefficient $GM$. In this long-range regime, however, the scalar contribution is approximately constant, $\Delta_0\simeq1/3$, and the orbital calibration gives
\begin{equation}
\mu_{\rm obs}\simeq\frac{4GM}{3a},
\qquad
U(r)\simeq-\frac{\mu_{\rm obs}}{r}.
\end{equation}
Thus, the factor $4/3a$ is absorbed into the gravitational parameter inferred from Newtonian orbital motion. However, notice that the corresponding spatial metric nevertheless differs from its general-relativistic form, since $\gamma_{\rm obs}\simeq1/2$. Hence, this case is not consistent with observations.

To suppress the scalar contribution to the metric potentials themselves, the common Yukawa correction must instead be negligible at the shortest scale outside the source at which the potential is tested.
Taking this minimum scale to be of order the source radius, \(r_{\rm min}\simeq
R_\star\), the relevant condition is
\begin{equation}
    \mathcal F(m_0R_\star)e^{-m_0R_\star}\ll1 .
\end{equation}
For a constant-density source, using the large-\(m_0R_\star\) behavior of the
form factor defined above, one obtains
\begin{equation}
    \mathcal F(m_0R_\star)e^{-m_0R_\star}
    \simeq
    \frac{3}{2(m_0R_\star)^2}.
\end{equation}
Thus the Yukawa correction is negligible throughout the exterior region
\(r\gtrsim R_\star\) whenever \(m_0R_\star\gg1\), and we have $\Delta_0\simeq0$, with an orbital calibration given by
\begin{equation}
    \mu_{\rm obs}\simeq\frac{GM}{a}.
\end{equation}
Thus, $\gamma_{\rm obs}\simeq1$, recovering the general-relativistic result.

\subsubsection{Hierarchical regime: \texorpdfstring{$m_-\ll m_+$}{m- << m+}}

A large difference in the masses may be relevant for the construction of successful cosmological
models, see for instance 
\cite{Sa2020}. Using the parametrization in
Eq.~\eqref{alpha-chi}, the weak-field potential can be written as
\begin{equation}
    U(r)=
    -\frac{GM}{ar}
    \left[
    1+
    \frac{\chi}{3}\mathcal F(m_-R_\star)e^{-m_-r}
    +
    \frac{1-\chi}{3}\mathcal F(m_+R_\star)e^{-m_+r}
    \right],
    \label{potential-hierarchical-masses}
\end{equation}
The hierarchy \(m_-\ll m_+\) does not, by itself, determine whether the scalar
corrections are negligible. From Eq.~\eqref{potential-hierarchical-masses}, the
two scalar contributions are controlled by
\begin{equation}
    \frac{\chi}{3}\mathcal F(m_-R_\star)e^{-m_-r},
    \qquad
    \frac{1-\chi}{3}\mathcal F(m_+R_\star)e^{-m_+r}.
\end{equation}
Therefore, recovering the Newtonian form of the potential requires the
contribution of both modes to be negligible over the region in which the
potential is tested. In the hierarchical regime, this can occur in different
ways.

\medskip
\noindent\textbf{\textit{(i) Heavy mode suppressed by a small effective weight.}}

The heavy scalar
may have a small effective weight in the potential, namely
\begin{equation}
    \chi \simeq 1 .
\end{equation}
In this case, the heavier-mode contribution is suppressed independently of the size of its Yukawa factor. However, since \(\chi\simeq1\), the Yukawa correction associated
with the lighter scalar mode has coefficient approximately \(1/3\). Hence, to recover the Newtonian form of the potential,
the lighter-mode Yukawa factor must also be negligible. Taking the smallest
scale outside the source to be of order the source radius, this requires
\begin{equation}
    \mathcal F(m_-R_\star)e^{-m_-R_\star}\ll1 .
\end{equation}
For a constant-density source, this condition is achieved when
\begin{equation}
    m_-R_\star\gg1 .
\end{equation}
Thus, we have $\Delta_0\simeq0$, with an orbital calibration given by
\begin{equation}
    \mu_{\rm obs}\simeq\frac{GM}{a},
\end{equation}
and $\gamma_{\rm obs}\simeq1$, recovering the general-relativistic result.

\medskip
\noindent\textbf{\textit{(ii) Heavy mode suppressed by finite range.}}

The heavier scalar may instead be suppressed by its finite range. Taking again
the smallest exterior scale to be of order \(R_\star\), the relevant condition
is
\begin{equation}
    \mathcal F(m_+R_\star)e^{-m_+R_\star}\ll1 .
    \label{heavy-mode-suppression}
\end{equation}
For a constant-density source, this condition is achieved when
\begin{equation}
    m_+R_\star\gg1 .
\end{equation}
Condition~\eqref{heavy-mode-suppression} ensures that the heavier-mode
contribution is negligible throughout the exterior region \(r\gtrsim R_\star\),
independently of the value of its weight \(1-\chi\). The potential then reduces
to
\begin{equation}
    U(r)\simeq
    -\frac{GM}{ar}
    \left[
    1+\frac{\chi}{3}\mathcal F(m_-R_\star)e^{-m_-r}
    \right].
    \label{potential-light-mode}
\end{equation}
This approximation does not require taking the formal limit
\(m_+\rightarrow\infty\); it only requires the heavier-mode contribution to be
negligible for \(r\gtrsim R_\star\).

There are then two relevant possibilities for the lighter mode.

\modecasecustom{ii-a}{Lighter mode long-ranged over the observational range.}

Let us first consider the case in which the lighter scalar is long-ranged over the source and over the observational region
outside the source. If \(\ell_{\rm max}\)
denotes the largest exterior distance at which the potential is probed, this
regime is characterized by
\begin{equation}
    m_-\ell_{\rm max}\ll1 .
\end{equation}
Then, for exterior radii $R_\star\lesssim r\leq \ell_{\rm max}$, we have that $\mathcal F(m_-R_\star)\simeq1$, and $e^{-m_-r}\simeq1 $. Therefore,
\begin{equation}
    \mathcal F(m_-R_\star)e^{-m_-r}\simeq1 .
\end{equation}
Equation~\eqref{potential-light-mode} then gives
\begin{equation}
    U(r)\simeq
    -\frac{GM}{ar}
    \left(1+\frac{\chi}{3}\right).
    \label{potential-long-range-light}
\end{equation}

In this regime the scalar contribution is approximately constant, $\Delta_0\simeq\chi/3$. Hence, the gravitational parameter inferred from the leading orbital dynamics and the corresponding PPN parameter are
\begin{equation}
\mu_{\rm obs}\simeq
\frac{GM}{a}\left(1+\frac{\chi}{3}\right),
\qquad
\gamma_{\rm obs}\simeq
\frac{1-\chi/3}{1+\chi/3}.
\end{equation}
The leading orbital potential is therefore Newtonian after calibration for any allowed value of $\chi$. Agreement of the 
{corresponding}
linear metric with its general-relativistic form, however, requires $\chi\ll1$ in this regime.

The limiting value
\begin{equation}
    U(r)\simeq -\frac{4GM}{3ar}
    \label{potential-four-thirds}
\end{equation}
is obtained when \(\chi\simeq1\). Using the definition of \(\chi\), this
corresponds to
\begin{equation}
    \mathcal B\simeq -\frac{1}{3m_+^2}.
\end{equation}
By contrast, when \(\chi\simeq0\), one has
\begin{equation}
    \mathcal B\simeq -\frac{1}{3m_-^2}.
\end{equation}
In this case, the lighter scalar mode has a small effective weight in
the potential, and the long-range correction to the Newtonian term is
suppressed even if \(m_-\) is very small. Therefore, if the heavier mode is suppressed by its finite range while the lighter mode remains long-ranged, values of $\chi$ such that $\chi\ll1$ suppress the departure of the calibrated modified spatial linear metric from the observations.

\modecasecustom{ii-b}{Lighter mode short-ranged over the observational range.}

Another possibility is that the lighter scalar mode is also suppressed by its
finite range. Taking the smallest exterior scale to be of order \(R_\star\),
the relevant condition is
\begin{equation}
    \mathcal F(m_-R_\star)e^{-m_-R_\star}\ll1 .
    \label{light-mode-suppression}
\end{equation}
As we saw before, for a constant-density source, this condition is achieved when \(m_-R_\star\gg1\).

Since the heavier mode is already suppressed by
Eq.~\eqref{heavy-mode-suppression}, both scalar corrections are negligible in
this case, yielding $\Delta_0\approx0$. Therefore, the Newtonian form is recovered independently of the
value of \(\chi\), provided that \(0<\chi<1\), and the orbital calibration
\begin{equation}
    \mu_{\rm obs}\simeq\frac{GM}{a}
\end{equation}
is adopted.

\medskip
\noindent\textbf{\textit{(iii) Both modes long-ranged over the observational range.}}

We finally consider the case in which, despite the hierarchy \(m_-\ll m_+\),
the heavier mode is still long-ranged over the observational region. If
\(\ell_{\rm max}\) denotes the largest exterior distance at which the potential
is probed, this regime is characterized by \(m_+\ell_{\rm max}\ll1\). Since
\(R_\star\leq \ell_{\rm max}\) and \(m_-\ll m_+\), this condition also implies
\(m_+R_\star\ll1\), \(m_-\ell_{\rm max}\ll1\), and \(m_-R_\star\ll1\).
Hence both scalar modes are long-ranged with respect to the source size and
throughout the observational region. For \(R_\star\lesssim r\leq
\ell_{\rm max}\), it follows that
\begin{equation}
    \mathcal F(m_iR_\star)e^{-m_ir}
    =
    1-m_ir
    +\mathcal{O}\!\left(m_i^2r^2,m_i^2R_\star^2\right).
    \label{Yukawa-long-range-expansion}
\end{equation}
Substitution into Eq.~\eqref{potential-hierarchical-masses} yields
\begin{equation}
    U(r)\simeq
    -\frac{GM}{ar}
    \left[
    1+\frac{1}{3}
    -\frac{r}{3}
    \left(
    \chi m_-+(1-\chi)m_+
    \right)
    +\mathcal{O}\!\left(m_+^2\ell_{\rm max}^2,m_+^2R_\star^2\right)
    \right].
    \label{potential-hierarchical-long-range-chi}
\end{equation}

Since \(0<\chi<1\), the weighted combination
\(\chi m_-+(1-\chi)m_+\) lies between \(m_-\) and \(m_+\). Therefore, under the
long-range condition \(m_+\ell_{\rm max}\ll1\), the
\(r\)-dependent correction inside the brackets is subleading with respect to
the constant scalar contribution \(1/3\). The leading \(1/r\) part of the
potential is consequently
\begin{equation}
    U(r)\simeq -\frac{4}{3}\frac{GM}{ar}.
\end{equation}

Thus, the factor $4/3a$ relative to the bare coefficient $GM$ is recovered independently of $\chi$. After orbital calibration, however, the leading potential takes the Newtonian form $U(r)\simeq-\mu_{\rm obs}/r$, with $\mu_{\rm obs}\simeq4GM/(3a)$, and $\gamma_{\rm obs}\simeq1/2$. Hence, this case is not consistent with observations.

The conditions under which the scalar contributions to the bare-normalized potential \eqref{potential-2} and $\gamma$ parameter \eqref{parameter-gamma-2} are suppressed can be summarized as follows, taking the minimum scale outside the source to be of order $R_\star$:
\begin{itemize}
    \item In the nearly degenerate regime, \(m_+\simeq m_-\simeq m_0\), the
    two scalar modes combine into a single Yukawa correction with total
    amplitude \(1/3\). Therefore, the scalar contributions are suppressed if
    \begin{equation}
        \mathcal F(m_0R_\star)e^{-m_0R_\star}\ll1 .
    \end{equation}
    For a constant-density source, this condition is satisfied when
    \(m_0R_\star\gg1\).

    \item In the hierarchical regime, \(m_-\ll m_+\), if the heavier mode is
    suppressed by its small effective weight, \(1-\chi\ll1\), then
    \(\chi\simeq1\), 
which entails that 
$\mathcal{B}\simeq -\frac{1}{3m_+^2}$. The Yukawa correction associated with the lighter scalar mode then has coefficient approximately \(1/3\). In this case, the corresponding scalar contribution is suppressed if
\begin{equation}
    \mathcal F(m_-R_\star)e^{-m_-R_\star}\ll1 .
\end{equation}
For a constant-density source, this requires that \(m_-R_\star\gg1\).

    \item In the hierarchical regime, \(m_-\ll m_+\), if the heavier mode is
    suppressed by its finite range, then
    \begin{equation}
        \mathcal F(m_+R_\star)e^{-m_+R_\star}\ll1 .
    \end{equation}
    For a constant-density source, this is achieved when
    \(m_+R_\star\gg1\). If the lighter mode is also short-ranged,
    \begin{equation}
        \mathcal F(m_-R_\star)e^{-m_-R_\star}\ll1 ,
    \end{equation}
    then the Newtonian form is recovered independently of \(\chi\). If,
    instead, the lighter mode is long-ranged over the observational region,
    \(m_-\ell_{\rm max}\ll1\), then the corresponding scalar contribution is suppressed only when
    its effective weight is small,
    \begin{equation}
        \chi\ll1,
        \qquad
        \mathcal B\simeq -\frac{1}{3m_-^2}.
    \end{equation}
 \item If both scalar modes are long-ranged over the source and over the
    observational region, their amplitudes add up to \(1/3\), independently of
    \(\chi\). In this case the leading potential becomes
    \begin{equation}
        U(r)\simeq -\frac{4}{3}\frac{GM}{ar},
    \end{equation}
    which differs from the potential normalized by the bare coefficient $GM$, but reduces to $U(r)\simeq-\mu_{\rm obs}/r$ after orbital calibration in this long-range regime.
\end{itemize}

Thus, suppression of the scalar contributions to the metric potentials imposes restrictions on their ranges and amplitudes. These restrictions are stronger than the requirement that an approximately constant scalar contribution be absorbed into the gravitational parameter governing the leading Newtonian orbital dynamics.

\subsection{Epicyclic radial frequency}\label{subsection 5.2}

Let us now consider the radial epicyclic frequency for timelike geodesic motion in the weak-field metric derived above. Restricting to the static and spherically symmetric case, we write the line element in Schwarzschild-like form as
\begin{equation}
    ds^2=-A(\tilde{r})\,dt^2+C(\tilde{r})\,d\tilde{r}^2+\tilde{r}^2d\Omega^2,
\end{equation}
where \(\tilde{r}\) denotes the areal radius.
Comparing this metric with Eq.~\eqref{newtonian-ds}, we find
\begin{align}
\tilde{r}
&=r\sqrt{1-2\gamma(r)U(r)},\\
A(\tilde{r})
&=1+2U\!\left(r(\tilde{r})\right),\\
C(\tilde{r})
&=\left.
\left[
1-\frac{r}{1-2\gamma(r)U(r)}
\frac{d}{dr}\bigl(\gamma(r)U(r)\bigr)
\right]^{-2}\right|_{r=r(\tilde{r})}.
\end{align}
The weak-field regime is defined by the requirement that the metric perturbations remain small, namely
\begin{equation}
    |U|\ll1, \qquad |\gamma U|\ll1.
\end{equation}
For the exterior solution considered here, these conditions are equivalent to demanding
\begin{equation}
    \frac{GM}{ar}\,\bigl|1\pm \Delta(r)\bigr|\ll 1,
\end{equation}
where, as before, 
\begin{equation}
\Delta(r)\equiv
\alpha_-\,\mathcal{F}(m_-R_\star)e^{-m_- r}
+\alpha_+\,\mathcal{F}(m_+R_\star)e^{-m_+ r}.
\end{equation}
In this regime, expanding to first order in \(\gamma U\) gives
\begin{equation}
    \tilde{r}\approx\left[1-\gamma(r)U(r)\right]r.
\end{equation}
Since \(\tilde{r}-r=\mathcal{O}(\gamma U)\), replacing \(r\) by \(\tilde{r}\) in the weak-field functions changes them only at an order higher than the first. Therefore, $
    U(r)\approx U(\tilde{r})$,
    $\gamma(r)\approx\gamma(\tilde{r})$, and
\begin{equation}
    A(\tilde{r})\approx1+2U(\tilde{r}).
    \label{113}
\end{equation}
For the radial metric coefficient, we obtain
\begin{equation}
    C(\tilde{r})\approx 1+2\tilde{r}\frac{d}{d\tilde{r}}\bigl[\gamma(\tilde{r})U(\tilde{r})\bigr].
    \label{115}
\end{equation}
Let us remark that, in a general case, an expansion of the metric coefficients including terms that are $\mathcal{O}(U_N^2)$ would be necessary in order to calculate the leading periapsis advance (see for instance \cite{Will2018}). We shall restrict below to a case in which the present calculation, which is linear in $U_N$, may be used.

Substituting Eqs.~\eqref{potential-2} and \eqref{parameter-gamma-2} into Eqs.~\eqref{113} and \eqref{115}, it follows that
\begin{equation}
    A(\tilde{r})=1-\frac{2GM}{a\tilde{r}}\left[
    1+\alpha_-\,\mathcal{F}(m_-R_\star)e^{-m_-\tilde{r}}
    +\alpha_+\,\mathcal{F}(m_+R_\star)e^{-m_+\tilde{r}}
    \right],
\end{equation}
and
\begin{equation}
    C(\tilde{r})=1+\frac{2GM}{a\tilde{r}}\left[
    1
    -\alpha_-(1+m_-\tilde{r})\,\mathcal{F}(m_-R_\star)e^{-m_-\tilde{r}}
    -\alpha_+(1+m_+\tilde{r})\,\mathcal{F}(m_+R_\star)e^{-m_+\tilde{r}}
    \right].
\end{equation}

It is convenient to define
\begin{equation}
P(\tilde{r})\equiv
\alpha_-(1+m_-\tilde{r})\,\mathcal{F}(m_-R_\star)e^{-m_-\tilde{r}}
+
\alpha_+(1+m_+\tilde{r})\,\mathcal{F}(m_+R_\star)e^{-m_+\tilde{r}},
\end{equation}
so that
\begin{equation}
\label{AC}
A(\tilde{r})=1-\frac{2GM}{a\tilde{r}}\bigl[1+\Delta(\tilde{r})\bigr],
\qquad
C(\tilde{r})=1+\frac{2GM}{a\tilde{r}}\bigl[1-P(\tilde{r})\bigr].
\end{equation}

The Lagrangian 
for the geodesics 
reads
\begin{equation}
   \mathcal{L}=\frac{1}{2}\left( -A(\tilde{r})\dot t^2 + C(\tilde{r})\dot{\tilde{r}}^2 +\tilde{r}^2\dot \theta^2 +\tilde{r}^2 \sin^2\theta\,\dot \varphi^2 \right),
\end{equation}
where overdots denote derivatives with respect to the proper time \(\tau\). Restricting to equatorial motion, \(\theta=\pi/2\), this reduces to
\begin{equation}
   \mathcal{L}=\frac{1}{2}\left( -A(\tilde{r})\dot t^2 + C(\tilde{r})\dot{\tilde{r}}^2  +\tilde{r}^2 \dot \varphi^2 \right).
   \label{lagrangian-equatorial}
\end{equation}
Since the Lagrangian is independent of \(t\) and \(\varphi\), the Euler--Lagrange equations imply the conserved quantities
\begin{equation}
    A(\tilde{r})\dot t=E,
    \qquad
    \tilde{r}^2 \dot \varphi=L,
\end{equation}
corresponding to the conserved energy and angular momentum per unit mass. Using the timelike normalization condition,
\begin{equation}
    g_{\mu\nu}\dot x^\mu \dot x^\nu
    =-A(\tilde{r})\dot t^2 + C(\tilde{r})\dot{\tilde{r}}^2 +\tilde{r}^2 \dot \varphi^2=-1,
    \label{ds-tipo-tiempo1}
\end{equation}
we obtain the radial equation
\begin{equation}
\frac{1}{2}\left(\frac{d\tilde{r}}{d\tau}\right)^2 +V_{\mathrm{eff}}(\tilde{r}; E, L)=0,
\label{geodesic-expand}
\end{equation}
with the effective potential given by
\begin{equation}
    V_{\mathrm{eff}}(\tilde{r};E,L)\equiv-\frac{E^2}{2A(\tilde{r})C(\tilde{r})}
    +\frac{1}{2C(\tilde{r})}\left(1+\frac{L^2}{\tilde{r}^2}\right).
    \label{Veff-radial}
\end{equation}

To compute the radial epicyclic frequency, a circular orbit with radius $r_0$ is perturbed (\(\tilde{r}=\tilde{r}_0+\delta\tilde{r}\)), while keeping \(E\) and \(L\) fixed. The oscillation frequency is \cite{Bambi2018}
\begin{equation}
    \Omega^2_{\text{rad}}=\frac{1}{\dot{t}^2}\frac{\partial^2 V_{\mathrm{eff}}}{\partial \tilde{r}^2}\Bigg|_{\tilde{r}=\tilde{r}_0}.
\end{equation}
Using the circular-orbit conditions
\begin{equation}
V_{\mathrm{eff}}(\tilde{r}_0;E,L)=0,
\qquad
\frac{\partial V_{\mathrm{eff}}}{\partial \tilde{r}}\Bigg|_{\tilde{r}=\tilde{r}_0}=0,
\end{equation}
it follows that
\begin{equation}
\frac{\partial^2V_{\mathrm{eff}}}{\partial \tilde{r}^2}
=\frac{L^2}{C(\tilde{r})\tilde{r}^3}
\left[
\frac{A''(\tilde{r})}{A'(\tilde{r})}
-2\frac{A'(\tilde{r})}{A(\tilde{r})}
+\frac{3}{\tilde{r}}
\right].
\end{equation}
The orbital angular velocity, given by
$$
    \omega\equiv\frac{d\varphi}{dt}=\frac{\dot\varphi}{\dot t}
    =\frac{L}{\dot t\,\tilde{r}^2},
$$
reduces, for circular motion, to the following expression:
\begin{equation}
    \omega_0^2\equiv\omega^2(\tilde{r}_0)=\frac{A'(\tilde{r}_0)}{2\tilde{r}_0}
    =
    \frac{GM}{a\tilde{r}_0^3}\bigl[1+P(\tilde{r}_0)\bigr]
    ,
\end{equation}
where we have used that $P(\tilde{r}_0)=\Delta(\tilde{r}_0)-\tilde{r}_0\Delta'(\tilde{r}_0)$.
\\
Therefore,
\begin{equation}
    \Omega^2_{\text{rad}}=\frac{\tilde{r}_0\omega^2_0}{C(\tilde{r}_0)}\left[
\frac{A''(\tilde{r}_0)}{A'(\tilde{r}_0)}
-2\frac{A'(\tilde{r}_0)}{A(\tilde{r}_0)}
+\frac{3}{\tilde{r}_0}
\right].
\label{epiciclyc-1}
\end{equation}
As a consequence of the results presented in Sec.~\ref{sec:mass-hierarchy},
we analyze next the changes in the periapsis advance,
\begin{equation}
    \vartheta \equiv
    2\pi\left(\frac{\omega_0}{\Omega_{\text{rad}}}-1\right),
\end{equation}
in the hierarchical regime
\(m_-\ll m_+\)
\footnote{A more general analysis, including limiting cases, is presented in the Appendix \ref{appb}.}. 
The phenomenologically relevant situation is the one in which the heavy mode is already suppressed at the shortest scale
outside the source,
which we take to be of order the source radius, while the light mode remains
long-ranged on the orbital scale, but suppressed by $\chi\ll 1$. This situation corresponds to \textit{(ii-a)} of Sec.~\ref{sec:mass-hierarchy}, where it was shown that $\Delta_0\simeq \chi/3$. In such a case, 
it follows from 
Eq.\eqref{calibrated-newtonian-potential} that the effect of the Yukawa-like terms lead only to a redefinition of the product $GM$, and the linear expansion of the metric coefficients in terms of the Newtonian potential can be safely used in the calculation of the periapsis precession \footnote{The condition $\tilde r_0\Delta'_0\ll 1$ is also needed, see below.}. 
\\
The conditions under which the case presented in \textit{(ii-a)} of Sec.~\ref{sec:mass-hierarchy}
is valid are
\begin{equation}
    \mathcal F(m_+R_\star)e^{-m_+R_\star}\ll1 .
   \end{equation}
and 
$m_-\ell_{\rm max}\ll 1$, which entails
\begin{equation}
    \mathcal F(m_-R_\star)e^{-m_- r}\simeq 1 
   \end{equation}
for $R_*\simeq r\leq \ell_{\rm max}$.
From
$$
-\tilde r_0\Delta'_0=
    \frac{\chi}{3}\tilde r_0
    m_-\mathcal F(m_-R_\star)e^{-m_-\tilde r_0}
    +
    \frac{1-\chi}{3}\tilde r_0 
    m_+\mathcal F(m_+R_\star)e^{-m_+\tilde r_0},
    $$
    and 
    the conditions given above, it follows that the inequality $|\tilde r_0\Delta'_0|\ll 1$
    is automatically satisfied.
Taking into account all these conditions, 
it follows from Eq.\eqref{AC} that
$$
A(\tilde r)=1-\frac{2\mu_{\rm obs}}{\tilde r},\;\;\;\;\;\;\;\;C(\tilde r) = 1+\frac{2\mu_{-}}{\tilde r}
,$$
where $\mu_{\rm obs}= \frac{GM}{a}\left(1+\frac{\chi}{3}\right)$, and 
$\mu_{-}\equiv \frac{GM}{a}\left(1-\frac{\chi}{3}\right)$. Hence, 
$$
\Omega_{\rm rad}^2=\omega_0^2\left[1-\frac{6GM}{\tilde r_0 a}\left(1+\frac{\chi}{9} \right)
\right],
$$
and the periapsis advance is
\begin{equation}
\label{padv}
\vartheta\simeq\frac{6\pi}{\tilde r_0} \frac{GM}{a}\left(1+\frac{\chi}{9}\right)=
\frac{2\pi}{\tilde r_0}\frac{(9+\chi)}{(3+\chi)}\mu_{\rm obs}.
\end{equation}
The latter reduces to the result obtained for the Schwarzschild metric, namely
$$
\vartheta_{GR} = \frac{6\pi}{\tilde r_0}GM,
$$
for $a=1$ and $\chi=0$.
Hence, we define
$$
\Delta\vartheta\equiv \vartheta - \vartheta_{GR}=\frac{6\pi}{\tilde r_0} GM\left[\frac{1}{a}\left(1+\frac{\chi}{9}\right)-1
\right],
$$
which is zero for $a=1$ and 
$\chi=0$.

We now apply our results to the Solar-System bounds on
anomalous planetary periapsis precession. Since 
the source is the Sun, we set \(R_\star=R_\odot\), and identify the
orbital radius \(\tilde r_0\) with the observed semimajor axis \(r_p\),
\begin{equation}
    \tilde r_0\simeq r_p .
\end{equation}
This identification is consistent with the weak-field approximation, since the
difference between \(\tilde r_0\) and \(r_p\) contributes only at higher
post-Newtonian order. 
In the following, we shall also assume that the independent weak-field constraint
\(a\simeq1\) is valid. Hence, Eq.~\eqref{padv} reduces to
\begin{equation}
\label{padvf}
    \Delta\vartheta = \frac{\pi\: r_s}{r_p}\frac{\chi}{3}.
\end{equation}
The orbital data used in the analysis are listed in
Table~\ref{tab:planetary-data}. Since the observed anomalous precession rates
are given in units of \({\rm mas\,yr^{-1}}\), the corresponding uncertainty
per orbit is
\begin{equation}
	\sigma_{\text{orb}}
	=
	\sigma_{\Delta\vartheta}\:T_p\:
	\frac{\pi}{648000000},
\end{equation}
where \(\sigma_{\Delta\vartheta}\) denotes the uncertainty in the anomalous precession rate in units of \({\rm mas\,yr^{-1}}\), and \(T_p\) is the orbital period in years.

\begin{table}[H]
\centering
\caption{Planetary orbital parameters and supplementary perihelion precession rates \cite{Berry2011}.}
\label{tab:planetary-data}
\begin{tabular}{lccc}
\hline\hline
Planet & 
Semimajor axis & 
Orbital period & 
Precession rate \\
& 
$r_p/10^{11}\,\mathrm{m}$ & 
$T_p/\mathrm{yr}$ & 
$\Delta\vartheta \pm \sigma_{\Delta\vartheta}\,(\mathrm{mas}\,\mathrm{yr}^{-1})$ \\
\hline
Mercury & 0.57909175 & 0.24084445 & $-0.040 \pm 0.050$ \\
Venus   & 1.0820893  & 0.61518257 & $0.24 \pm 0.33$ \\
Earth   & 1.4959789  & 0.99997862 & $0.06 \pm 0.07$ \\
Mars    & 2.2793664  & 1.88071105 & $-0.07 \pm 0.07$ \\
Jupiter & 7.7841202  & 11.85652502 & $0.67 \pm 0.93$ \\
Saturn  & 14.267254 & 29.42351935 & $-0.10 \pm 0.15$ \\
Uranus  & 28.709722 & 83.74740682 & $-38.9 \pm 39.0$ \\
Neptune & 44.982529 & 163.7232045 & $-44.4 \pm 54.0$ \\
Pluto   & 59.063762 & 248.0208 & $28.4 \pm 45.1$ \\
\hline\hline
\end{tabular}
\end{table}
Requiring the predicted anomalous precession to lie within the observational
uncertainty,
\begin{equation}
    |\Delta\vartheta|\le\sigma_{\text{orb}},
\end{equation}
and using Eq.~\eqref{padvf} and $a\simeq 1$, we obtain
\begin{equation}
    \chi
    \le
    \chi_{\max}
    \equiv
    \frac{r_p}{r_s}\frac{\sigma_{\Delta\vartheta}T_p}{216000000}.
    \label{chi-bound-planet}
\end{equation}
Only values \(\chi_{\max}<1\) provide nontrivial bounds, since the allowed
sector already requires \(0<\chi<1\).

It is important to point out that the constraint on $\chi$ are approximate order-of-magnitude sensitivity estimates obtained
under fixed-ephemeris and one-parameter assumptions.

Using Eq.~\eqref{B-chi}, the same bound can be written as a constraint on
\(\mathcal B\):
\begin{equation}
    -\frac{1}{3m_-^2}
    <
    \mathcal B
    \le
    -\frac{1}{3m_-^2}
    +
    \chi_{\max}
    \left(
    \frac{1}{3m_-^2}
    -
    \frac{1}{3m_+^2}
    \right).
    \label{B-bound-planet}
\end{equation}
For \(m_-\ll m_+\), this becomes
\begin{equation}
    -\frac{1}{3m_-^2}
    <
    \mathcal B
    \lesssim
    -\frac{1-\chi_{\max}}{3m_-^2}.
    \label{B-bound-planet-hierarchical}
\end{equation}

\begin{table}[H]
\caption{Bounds on \(\chi\) from anomalous periapsis precession in the viable
hierarchical regime. We assume a suppressed heavy mode,
\(m_+ r_p\gg1\), a long-ranged light mode,
\(m_-r_p\ll1\), \(m_-R_\odot\ll1\), and \(a\simeq1\).}
\label{tab:chi_bounds_viable_hierarchy}
\centering
\small
\setlength{\tabcolsep}{5pt}
\begin{tabular}{lc}
\hline\hline
Planet
& \(\chi_{\max}\) \\
\hline
Mercury & \(1.08\times10^{-3}\) \\
Venus   & \(3.39\times10^{-2}\) \\
Earth   & \(1.62\times10^{-2}\) \\
Mars    & \(4.63\times10^{-2}\) \\
Jupiter & \(13.25\) \\
Saturn  & \(9.72\) \\
Uranus  & \(14470.70\) \\
Neptune & \(61372.40\) \\
Pluto   & \(101956.00\) \\
\hline\hline
\end{tabular}
\end{table}

The bounds in Table~\ref{tab:chi_bounds_viable_hierarchy} show that only the
inner planets provide nontrivial restrictions on \(\chi\). For Jupiter and the
outer planets one finds \(\chi_{\max}>1\), so these entries do not further
restrict the allowed values of $\chi$. Among the allowed bounds, the
constraint that follows from 
Mercury is the strongest. At \(1\sigma\), it yields
\begin{equation}
    \chi\lesssim 1.08\times10^{-3}.
\end{equation}
Thus, for the strongest planetary constraint, the allowed region lies well
inside the small-\(\chi\) regime. Therefore, once the heavy mode is suppressed
over the Solar-System exterior region, periapsis constraints require the
remaining long-ranged light mode to have a very small effective weight in the
Solar-System potential.  This means that the
parameter 
\(\mathcal B\) 
must be correspondingly close to the lower endpoint
\(-1/(3m_-^2)\) of the ghost- and tachyon-free interval given in Eq.
\eqref{B-interval}.

\section{Concluding remarks}
\label{concl}

In this work we analyzed the weak-field sector of generalized hybrid metric-Palatini gravity, defined by a general function \(f(R,\mathcal{R})\) of the metric and Palatini curvature scalars.
Starting from the 
field equations 
for the 
metric and the connection 
{in the original representation (that is, without transforming the theory to the Jordan or Einstein frame)}, and assuming vanishing torsion, the theory was linearized around a Minkowski background with vanishing effective cosmological constant. At linear order, the independent connection can be expressed in terms of a conformally-related auxiliary metric, allowing the
{perturbed}
field equations to be written solely in terms of the metric perturbation and the first-order curvature scalars,
{without assuming that $f^{(0)}_{,\mathcal{R}\mathcal{R}}=0$.}

{We showed, as expected, that} the linearized theory separates into a massless spin-2 sector and a scalar sector. The spin-2 perturbation satisfies the standard wave equation but, differently from GR, the gravitational coupling is rescaled according to
\[
\kappa_{\rm eff}
=
\frac{\kappa}{f^{(0)}_{,R}+f^{(0)}_{,\mathcal{R}}}.
\]
The scalar part of the trace equation factorizes into two Klein--Gordon operators, revealing two scalar degrees of freedom with masses \(m_+\) and \(m_-\), which are given in terms of the 
first and second order derivatives of $f$ evaluated at the background. This provides a direct perturbative interpretation of the additional degrees of freedom present in the generalized hybrid theory and makes explicit the conditions under which the general-relativistic limit is recovered. Degenerate cases, in which the masses coincide, one of the scalar modes is absent, and both modes are non-dynamical, have also been examined.

We also studied the propagator of the linearized theory using the Barnes--Rivers spin-projector formalism \cite{Biswas2011,Koivisto2013}. The propagator contains the usual massless spin-2 pole and two additional scalar poles associated with the masses \(m_\pm\). Requiring these scalar modes to be free from tachyonic and ghost instabilities leads to algebraic restrictions on the derivatives of \(f(R,\mathcal{R})\) evaluated on the Minkowski background. For the generic nondegenerate two-scalar branch, these restrictions include the positivity of the effective gravitational coupling, the absence of tachyonic scalar masses, the absence of negative-residue scalar poles, and the exclusion of the vanishing-residue degeneracy. These conditions select the viable linearized sector of the theory, and generalize those obtained in \cite{Bombacigno2019}.

The weak-field phenomenology was obtained by solving the linearized equations for a static spherical source, extending previous analyses of the weak-field regime of generalized hybrid metric-Palatini gravity, 
{which were restricted to point-like masses}
\cite{Rosa2021,Bombacigno2019}. The Newtonian potential receives one Yukawa-like 
correction from each scalar mode, with amplitudes determined by the residues of the scalar poles. For an extended source, these corrections are weighted by form factors that depend on the density profile and on the combinations \(m_\pm R_\star\). In the point-particle limit, the form factors reduce to unity and the result becomes the known two-Yukawa correction to the Newtonian potential, each with an amplitude. As a consistency check, 
we showed that
the known result for metric $f(R)$ and hybrid theories \(f(R,\mathcal{R})=R+\xi(\mathcal{R})\) are recovered in the appropriate limit \cite{Capo2015,Harko2011}.

The corresponding weak-field metric contains Yukawa contributions in the metric potentials, leading to an effective post-Newtonian function \(\gamma(r)\) that depends on the radial coordinate. Therefore, the usual Solar-System bounds on a constant PPN parameter \(\gamma\) cannot be imposed directly. Instead, observational constraints must be derived from the metric itself, for example by studying light propagation or orbital motion in the modified weak-field geometry
\cite{Will2014,Toniato2021,Huang2024,Berry2011}
. In particular, when $\Delta(r)$ varies with radius over the relevant observational region, we have shown that  light propagation is governed by the effective combination \(\gamma_\Sigma=2/a-1\), with \(a=f^{(0)}_{,R}+f^{(0)}_{,\mathcal{R}}\). This distinction is important for theories in which the post-Newtonian corrections contain Yukawa factors and source-dependent form factors.

We further analyzed the phenomenology in different scalar-mass regimes and identified the conditions under which the Newtonian limit is recovered. In the nearly degenerate regime, \(m_+\simeq m_-\), the two scalar modes combine into a single effective Yukawa correction, so the Newtonian form is recovered when this common scalar contribution is sufficiently short-ranged. In the hierarchical regime, \(m_-\ll m_+\), a viable Newtonian limit is obtained when the heavy mode is suppressed and the remaining light mode is either short-ranged or has a sufficiently small effective weight, \(\chi\ll1\). Thus, the recovery of the Newtonian limit follows from the scalar mass spectrum and residue structure of the theory, without invoking an additional screening mechanism such as chameleon screening in metric \(f(R)\) gravity
\cite{Burrage2017}.

We also derived the radial epicyclic frequency for timelike geodesics in the weak-field metric and used it to obtain the corresponding anomalous periapsis advance \cite{Bambi2018,Berry2011}.{
In the case $m_-\ll m_+$ and considering 
that the heavier mode is suppressed by its finite range, the periapsis-precession analysis gives direct bounds on \(\chi\),
which characterizes the remaining 
long-range light scalar contribution}. In the hierarchical sub-regime 
considered, 
and
assuming \(a\simeq1\), the strongest constraint is obtained from Mercury, using the planetary periapsis-precession data quoted in Ref.~\cite{Berry2011}:
\[
\chi\lesssim 1.08\times10^{-3}.
\]
Thus, if the light scalar is long-ranged on planetary scales, its effective contribution to the Solar-System potential must be strongly suppressed. Equivalently, the parameter \(\mathcal{B}\) must lie close to the lower endpoint of the ghost- and tachyon-free interval.
{As a byproduct, we have obtained the expression of the radial epicyclic frequency for timelike geodesics in the weak-field metric for theories of the type
$f=R+\xi(\mathcal{R})$.}

To summarize, our results provide a unified weak-field description of generalized hybrid metric-Palatini gravity. They clarify the propagating content of the theory, identify the linear stability and nondegeneracy requirements of the scalar sector, present the modified Newtonian potential for extended sources, and connect the viable parameter space with Solar-System observables. Future extensions of our work  include a dedicated treatment of degenerate scalar sectors, the calculation of the 
periapsis advance in cases in which the second-order expansion in the Newtonian potential is needed,
the calculation of the modified equations 
of motion in $f(R,\mathcal{R})$ theories in a detailed planetary ephemeris model (and consequence fit with observational data), 
a full light-deflection and time-delay analysis based on the radially dependent weak-field metric, and the application of the stability and Solar-System constraints to specific functional forms of \(f(R,\mathcal{R})\) motivated by cosmology.


\section{Acknowledgments}

G.M. acknowledges financial support from the Coordena\c{c}\~ao de Aperfei\c{c}oamento de Pessoal de N\'ivel Superior -- Brasil (CAPES) -- Finance Code 001.
SEPB acknowledges support from of CNPq, grant PQ-C 300241/2025-9. The authors would like to thank Marcelo Santos Guimar\~aes for a helpful discussion.

\appendix

\section{Effective post-Newtonian parameter for light propagation}
\label{appendix}

We now present the calculation that identifies the effective parameter controlling light propagation. Consider the weak-field line element
\begin{equation}
ds_N^2=-(1+2\sigma_e \mathcal{U})dt^2
+(1-2\gamma_e \mathcal{U})\delta_{ij}dx^i dx^j,
\label{metric_pn_eff}
\end{equation}
where \(\sigma_e\) and \(\gamma_e\) are, in general, functions of position, and
\begin{equation}
\mathcal{U}(r)=-\frac{GM}{r}
\label{Newtonian_potential}
\end{equation}
is the Newtonian potential \footnote{ This form is equivalent, up to conventions, to the
one adopted in \cite{Toniato2021}}. Writing
$$
g^N_{\mu\nu}=\eta_{\mu\nu}+h^N_{\mu\nu},
$$
it follows that
\begin{equation}
h^N_{00}=-2\sigma_e \mathcal{U}, \qquad
h^N_{ij}=-2\gamma_e \mathcal{U}\,\delta_{ij}.
\label{hN_components}
\end{equation}
Therefore,
\begin{equation}
\sigma_e=-\frac{h^N_{00}}{2\mathcal{U}},
\qquad
\gamma_e=-\frac{h^N_{ii}}{2\mathcal{U}},
\label{alpha_gamma_from_h}
\end{equation}
with no sum over \(i\).
Using Eqs.~\eqref{h^N_00} and \eqref{h^N_ij}, it follows that
\begin{equation}
\sigma_e=
\frac{1}{a}\left[
1+\alpha_-\,\mathcal{F}(m_-R_\star)e^{-m_- r}
+\alpha_+\,\mathcal{F}(m_+R_\star)e^{-m_+ r}
\right],
\label{sigma_e}
\end{equation}
and
\begin{equation}
\gamma_e=
\frac{1}{a}\left[
1-\alpha_-\,\mathcal{F}(m_-R_\star)e^{-m_- r}
-\alpha_+\,\mathcal{F}(m_+R_\star)e^{-m_+ r}
\right].
\label{gamma_e}
\end{equation}

To obtain the combination of parameters related to observations, determined by 
equation obeyed by null geodesics, we closely follow the derivation of the post-Newtonian
light-propagation equations presented in Refs.~\cite{Toniato2021,Berry2011},
adapted here to the metric in Eq.\eqref{metric_pn_eff}.
Let \(x^\mu(t)\) denote the photon trajectory, and define \(v^\mu=dx^\mu/dt=(1,v^i)\). Since the photon worldline is null, it satisfies
\begin{equation}
g^N_{\mu\nu}v^\mu v^\nu=0.
\label{null_condition_1}
\end{equation}
Substituting Eq.~\eqref{metric_pn_eff} into Eq.~\eqref{null_condition_1}, one obtains
\begin{equation}
-(1+2\sigma_e\mathcal{U})
+(1-2\gamma_e\mathcal{U})\,\delta_{ij}v^i v^j=0.
\end{equation}
To first post-Newtonian order this gives
\begin{equation}
\delta_{ij}v^i v^j
\approx1+2(\sigma_e+\gamma_e)\mathcal{U}.
\label{v_squared}
\end{equation}
Introducing a unit vector \(n^i\), with
\(\delta_{ij}n^i n^j=1\), we may write
\begin{equation}
v^i\approx\left[1+(\sigma_e+\gamma_e)\mathcal{U}\right]n^i.
\label{vi_expansion}
\end{equation}
Let us now consider the geodesic equation,
\begin{equation}
\frac{d^2x^\mu}{d\lambda^2}
+\Gamma^\mu_{\alpha\beta}
\frac{dx^\alpha}{d\lambda}\frac{dx^\beta}{d\lambda}=0,
\label{geodesic_affine}
\end{equation}
where \(\lambda\)
is an affine parameter. Rewriting the spatial components in terms
of the coordinate time \(t\), 
\begin{equation}
\frac{dv^i}{dt}
=
-\left(
\Gamma^i_{\alpha\beta}
-v^i\Gamma^0_{\alpha\beta}
\right)v^\alpha v^\beta.
\label{geodesic_tparam}
\end{equation}
For the metric \eqref{metric_pn_eff}, the leading Christoffel symbols at
post-Newtonian order are
\begin{equation}
\Gamma^i_{00}\approx\partial^i\left(\sigma_e\,\mathcal{U}\right),
\qquad
\Gamma^0_{0i}\approx\partial_i\left(\sigma_e\,\mathcal{U}\right),
\qquad
\Gamma^i_{jk}
\approx\left(\delta_{jk}\,\partial^i\left(\gamma_e\,\mathcal{U}\right)
-\delta^i{}_j\,\partial_k \left(\gamma_e\,\mathcal{U}\right)
-\delta^i{}_k\,\partial_j\left(\gamma_e\,\mathcal{U}\right)\right).
\label{Christoffel_eff}
\end{equation}
Substituting Eqs.~\eqref{vi_expansion} and \eqref{Christoffel_eff} into
Eq.~\eqref{geodesic_tparam}, and retaining only terms through first
post-Newtonian order, 
\begin{equation}
\frac{dn^i}{dt}
=-
\left(\delta^{ij}-n^i n^j\right)
\partial_j\!\left[(\sigma_e+\gamma_e)\mathcal U\right].
\label{dnidt}
\end{equation}
Comparing Eq.~\eqref{dnidt} with the standard PPN light-propagation
equation, it is seen that the projected-gradient structure is unchanged, while
the factor \(1+\gamma\) multiplying \(\mathcal U\) is replaced by
\(\sigma_e+\gamma_e\). Thus, the effective post-Newtonian parameter
governing null geodesics in this limit is
\begin{equation}
\gamma_\Sigma \equiv \sigma_e+\gamma_e-1,
\label{gamma_sigma_def}
\end{equation}
which is the combination identified in
Refs.~\cite{Toniato2021,Berry2011}.
\\
Using Eqs.~\eqref{sigma_e} and \eqref{gamma_e}, the Yukawa contributions cancel
exactly in the sum, so that
\begin{equation}
\sigma_e+\gamma_e=\frac{2}{a},
\end{equation}
and therefore
\begin{equation}
\gamma_\Sigma=\frac{2}{a}-1.
\label{gamma_sigma_final}
\end{equation}
The general-relativistic value is recovered for
\(a=1\).
It follows that, although \(\gamma_e(r)\) is position dependent, 
Solar-System light-deflection and time-delay experiments could be used to constrain the constant combination
$a=f_{,R}^{(0)}+f_{,\mathcal{R}}^{(0)}$.

If $\Delta(r)\simeq\text{constant}$  over the relevant observational region, the orbital calibration must be applied rather than using the bare coefficient $GM$. Then, we have
\begin{equation}
\frac{dn^i}{dt}
=-
\left(\delta^{ij}-n^i n^j\right)
\partial_j\!\left[\frac{(\sigma_e+\gamma_e)a}{(1+\Delta_0)}\left(-\frac{\mu_{\rm obs}}{r}\right)\right].
\label{dnidt_final}
\end{equation}

Comparing Eq.~\eqref{dnidt_final} with the standard PPN light-propagation
equation, it is seen that the projected-gradient structure is unchanged, while
the factor $1+\gamma$ multiplying $-\mu_{\rm obs}/r$ is replaced by
$(\sigma_e+\gamma_e)a/(1+\Delta_0)$.

Thus, the effective post-Newtonian parameter
governing null geodesics in this limit is
\begin{equation}
\gamma_\Sigma=\frac{(\sigma_e+\gamma_e)a}{(1+\Delta_0)}-1=\frac{1-\Delta_0}{1+\Delta_0}.
\end{equation}
Therefore, we recover the result for $\gamma_{\rm obs}$, shown in Eq.~\eqref{gamma-observed}.

The general-relativistic value is recovered when $\Delta_0\approx0$.

\section{Reduction of the expression for the periapsis advance to known cases}
\label{appb}
This appendix continues the framework established in Subsection~\ref{subsection 5.2}, evaluating the case where $\Delta(r)$ is no longer treated as constant. We examine key physical limits, specifically verifying the recovery of General Relativity and $f(R)$ gravity. For this purpose, it is useful to define
\begin{equation}
Q(\tilde{r})\equiv
\alpha_-m_-^2\,\mathcal{F}(m_-R_\star)e^{-m_-\tilde{r}}
+
\alpha_+m_+^2\,\mathcal{F}(m_+ R_\star)e^{-m_+\tilde{r}}.
\end{equation}
Then, expanding Eq.~\eqref{epiciclyc-1} consistently to first order in the weak-field parameter \(GM/(a\tilde{r})\) and using the Schwarzschild radius \(r_s\equiv 2GM\), 
\begin{equation}
\label{omegarad}
    \Omega^2_{\rm rad}
    =
    \omega_0^2
    \left[
    1-\frac{3r_s}{a\tilde{r}_0}
    -\zeta(\tilde{r}_0)
    \right],
\end{equation}
where
\begin{equation}
\zeta(\tilde{r}_0)
=\frac{r_s}{a\tilde{r}_0}P(\tilde{r}_0)+\frac{\tilde{r}_0^2 Q(\tilde{r}_0)}{1+P(\tilde{r}_0)}\left[
1-\frac{r_s}{a\,\tilde{r}_0}\,
[1-P(\tilde{r}_0)]\right].
\label{zeta-definition}
\end{equation}
Thus, \(\zeta(\tilde{r}_0)\) parametrizes the Yukawa-dependent contribution
to the radial epicyclic frequency. For \(a\neq1\), the total deviation from
the general-relativistic result also contains the rescaling of the Newtonian coupling
through \(a\), as shown below.

In the general-relativistic limit, obtained by taking \(m_\pm\to\infty\), the full Yukawa factors vanish
outside the source,
\begin{equation}
\mathcal{F}(m_\pm R_\star)e^{-m_\pm \tilde{r}}\longrightarrow 0,
\qquad
\tilde{r}\ge R_\star,
\end{equation}
so that \(P(\tilde{r})\to 0\) and \(Q(\tilde{r})\to 0\). Consequently, \(\zeta(\tilde{r}_0)\to 0\), and the radial epicyclic frequency reduces to
\begin{equation}
\Omega_{\rm rad}^2=\omega_0^2\left(1-\frac{3r_s}{a\tilde{r}_0}\right).
\end{equation}
The standard general-relativistic weak-field expression is then recovered for \(a=1\).
\\
The result for metric \(f(R)\) gravity follows from imposing  \(m_+\rightarrow\infty\), \(\alpha_-\rightarrow1/3\), and \(a=1\). Assuming \(m_- R_\star\ll1\), we obtain
\begin{equation}
    P(\tilde{r})\approx\frac{1}{3}(1+m_- \tilde{r})\,e^{-m_-\,\tilde{r}},\qquad Q(\tilde{r})\approx\frac{1}{3}m_-^2\,e^{-m_-\,\tilde{r}},
\end{equation}
so that \(\zeta(\tilde{r}_0)\) reduces to
\begin{equation}
\zeta(\tilde{r}_0)
=
\frac{1}{3}\frac{r_s}{\tilde{r}_0}(1+m_-\,\tilde{r}_0)\,e^{-m_-\,\tilde{r}_0}+\frac{\tilde{r}_0^2\,m_-^2\,e^{-m_-\, \tilde{r}_0}}{3+(1+m_-\,\tilde{r}_0)\,e^{-m_-\, \tilde{r}_0}}\left[1-\frac{r_s}{\tilde{r}_0}+\frac{1}{3}\frac{r_s}{\tilde{r}_0}(1+m_-\,\tilde{r}_0)\,e^{-m_-\,\tilde{r}_0}\right],
\label{zeta-fR}
\end{equation}
which coincides with the result presented in Ref.~\cite{Berry2011}.
\\
To obtain the result
for the hybrid theory \(f(R,\mathcal{R})=R+\xi(\mathcal{R})\), 
we impose that
\(m_+\rightarrow\infty\), \(\alpha_-\rightarrow-\xi^{(0)}_{,\mathcal{R}}/3\), and \(a\rightarrow1+\xi^{(0)}_{,\mathcal{R}}\). Assuming \(m_- R_\star\ll1\), it follows that
\begin{equation}
    P(\tilde{r})\approx-\frac{\xi^{(0)}_{,\mathcal{R}}}{3}(1+m_- \tilde{r})\,e^{-m_-\,\tilde{r}},\qquad Q(\tilde{r})\approx-\frac{\xi^{(0)}_{,\mathcal{R}}}{3}m_-^2\,e^{-m_-\,\tilde{r}},
\end{equation}
so that \(\zeta(\tilde{r}_0)\) reduces to
\begin{equation}
\zeta(\tilde{r}_0)
=
-\frac{\xi^{(0)}_{,\mathcal{R}}}{3}\frac{r_s}{a\,\tilde{r}_0}(1+m_-\,\tilde{r}_0)\,e^{-m_-\,\tilde{r}_0}-\frac{\xi^{(0)}_{,\mathcal{R}}\tilde{r}_0^2\,m_-^2\,e^{-m_-\, \tilde{r}_0}}{3-\xi^{(0)}_{,\mathcal{R}}(1+m_-\,\tilde{r}_0)\,e^{-m_-\, \tilde{r}_0}}\left[1-\frac{r_s}{a\,\tilde{r}_0}-\frac{\xi^{(0)}_{,\mathcal{R}}}{3}\frac{r_s}{a\,\tilde{r}_0}(1+m_-\,\tilde{r}_0)\,e^{-m_-\,\tilde{r}_0}\right].
\label{zeta-fX}
\end{equation}

For small deviations, the periapsis advance per orbit is
\begin{equation}
    \vartheta =
    2\pi\left(\frac{\omega}{\Omega_{\text{rad}}}-1\right).
\end{equation}
Using Eq.\eqref{omegarad} and expanding to first order, it follows that
\begin{equation}
    \vartheta
    \simeq
    \pi\left[
    \frac{3r_s}{a\tilde r_0}
    +\zeta(\tilde r_0)
    \right].
\end{equation}
Therefore, the anomalous periapsis advance relative to the GR prediction is
\begin{equation}
    \Delta\vartheta
    \equiv
    \vartheta-\vartheta_{\rm GR}
    \simeq
    \pi\left[
    \frac{3r_s}{\tilde r_0}\left(\frac{1}{a}-1\right)
    +\zeta(\tilde r_0)
    \right].
    \label{delta-vartheta-general}
\end{equation}
This expression shows that, in general, deviations from the general-relativistic periapsis advance receive two contributions: one from the rescaling of the Newtonian coupling through \(a\), and another from the Yukawa correction encoded
in \(\zeta(\tilde r_0)\).

\bibliography{biblio}

\end{document}